\documentstyle[sprocl]{article}
\newcommand{\ket}[1]{|{#1}\rangle}

\input{psfig}
\def\be{\begin{equation}}
\def\ee{\end{equation}}
\def\bea{\begin{eqnarray}}
\def\eea{\end{eqnarray}}
\begin{document}

\title{FAULT-TOLERANT QUANTUM COMPUTATION}

\author{ JOHN PRESKILL }

\address{California Institute of Technology, Pasadena, CA 91125, USA}


\maketitle\abstracts{
The discovery of {\it quantum error correction} has greatly improved the
long-term prospects for quantum computing technology.  Encoded quantum
information can be
protected from errors that arise due to uncontrolled interactions with the
environment, or due to imperfect implementations of quantum logical operations.
 Recovery from errors can work effectively even if occasional
mistakes occur during the recovery procedure.  Furthermore, encoded quantum
information can be processed without serious propagation of errors.
In principle, an arbitrarily long quantum computation can be performed
reliably,
provided that the average probability of error per quantum gate is less than a
certain critical value, the {\it accuracy threshold}. It may be possible to
incorporate intrinsic fault tolerance into the design of quantum computing
hardware, perhaps by invoking topological Aharonov-Bohm interactions to process
quantum information.}

\section{The need for fault tolerance}

Quantum computers appear to be capable, at least in principle, of solving
certain problems far faster than any conceivable classical
computer.\cite{feynman}$^{\!-\,}$\cite{shor_a} In practice, though, quantum
computing technology is still in its infancy.  While a practical and useful
quantum computer may eventually be constructed, we cannot clearly envision at
present what the hardware of that machine will be like. Nevertheless, we can be
quite confident that any practical quantum computer will
incorporate some type of error correction into its operation.  Quantum
computers are far more susceptible to making errors than conventional digital
computers, and some method of controlling and correcting those errors will be
needed to prevent a quantum computer from crashing.

The most formidable enemy of the quantum computer is {\it
decoherence}.\cite{landauer_a}$^{\!-\,}$\cite{haroche} We know how to prepare a
quantum state of a cat that is a superposition of a dead cat and a live cat,
but we never observe such macroscopic superpositions because they are very
unstable.  No real cat can be perfectly isolated from its environment.  The
environment measures the cat, in effect, immediately projecting it onto a state
that is completely alive or completely dead.\cite{zurek}  A quantum computer
may not be as complex as a cat, but it is a complicated quantum system, and
like a cat it inevitably interacts with the environment.  The information
stored in the computer decays, resulting in errors and the failure of the
computation.  Can we protect a quantum computer from the debilitating effects
of decoherence?

And decoherence is not our only
enemy.\cite{landauer_a}$^{\!-\,}$\cite{landauer_c} Even if we {\it were} able
to achieve excellent isolation of our computer from the environment, we could
not expect to execute quantum logic gates with perfect accuracy.  As with an
analog classical computer, the errors in the quantum gates form a continuum.
Small errors in the gates can accumulate over the course of a computation,
eventually causing failure, and it is not obvious how to correct these small
errors. Can we prevent the catastrophic accumulation of the small gate errors?

The future prospects for quantum computing received a tremendous boost from the
discovery\cite{shor_b}$^{\!-\,}$\cite{steane_b} that quantum error correction
is really possible in principle (see the preceding chapter by A. Steane).  But
this discovery in itself is not sufficient to ensure that a noisy quantum
computer can perform reliably.  To carry out a quantum error-correction
protocol, we must first encode the quantum information we want to protect, and
then repeatedly perform recovery operations that reverse the errors that
accumulate.
But encoding and recovery are themselves complex quantum
computations, and errors will inevitably occur while we perform these
operations.  Thus, we need to find methods for recovering from errors that are
sufficiently robust to succeed with high
reliability even when we make some errors during the recovery step.

Furthermore, to operate a quantum computer, we must do more than just {\it
store}
quantum information; we must {\it process} the information. We need to be able
to perform quantum gates, in which two or more encoded qubits come together and
interact with one another.  If an error occurs in one qubit, and then that
qubit interacts with another through the operation of a quantum gate, the error
is likely to spread to the second qubit.  We must design our gates to minimize
the propagation of error.

Incorporating quantum error correction will surely complicate the operation of
a quantum computer.  To establish the redundancy needed to protect against
errors, the number of elementary qubits will have to rise.  Performing gates on
encoded information, and inserting periodic error-recovery steps, will slow the
computation down.  Because of this necessary increase in the complexity of the
device, it is not {\it a priori} obvious that error correction will really
improve its performance.

A device that works effectively even when its elementary components are
imperfect is said to be {\it fault tolerant}.  This chapter is devoted to the
theory of fault-tolerant quantum computation.  We will address the issues and
questions summarized above.

In fact, similar issues also arise in the theory of fault-tolerant {\it
classical} computation.  Because existing silicon-based  circuitry is so
remarkably reliable, fault-tolerance is not essential to the operation of
modern digital computers. Even so, the study of fault-tolerant classical
computing has a distinguished history.  In 1952, Von Neumann~\cite{vonneumann}
suggested improving the reliability of a circuit with noisy gates by executing
each gate many times, and using majority voting.  He concluded that if the gate
failures are statistically independent, and the probability of failure per gate
is small enough, then any computation can be performed with reasonable
reliability.  One shortcoming of Von Neumann's analysis was that he assumed
perfect transmission of bits through the ``wires'' connecting the
gates.\footnote{This problem is serious because Von Neumann's circuits cannot
be realized in three-dimensional space with wires of bounded length; one would
expect the probability of a transmission error to approach unity as the wire
becomes arbitrarily long.}   Going beyond this assumption proved difficult, but
was eventually achieved in 1983 by G\'acs,\cite{gacs} who described a universal
cellular automaton with a hierarchical organization that can be maintained by
local operations in the presence of noise, without any need for direct nonlocal
communication among the components.

It is an interesting question whether a quantum system can similarly maintain a
complex hierarchical structure, but we will not be so ambitious as to address
this question here.  Because we are interested in the limitations imposed by
noise on the processing of {\it quantum} information, we will classify our
gates into classical and quantum, and we will assume that the classical gates
can be executed with perfect accuracy and as quickly as
necessary.\footnote{However, when we consider error recovery with quantum codes
of arbitrarily large block size, we {\it will} insist that the amount of
classical processing to be performed remains bounded by a polynomial in the
block size.}  This assumption will be well justified as long as the clock speed
and accuracy of our classical computer far exceed those of the quantum
computer.

After reviewing the features of a particular quantum error-correcting code
(Steane's 7-qubit code\cite{steane_b}) in \S2, we assemble the ingredients of
fault-tolerant recovery in \S3.  Errors that occur during recovery can further
damage the encoded quantum information; hence recovery must be implemented
carefully to be effective.  Ancilla qubits are used to measure an {\it error
syndrome} that diagnoses the errors in the encoded data block, and we must
minimize the propagation of errors from the ancilla to the data.  Methods for
controlling error propagation during recovery (proposed by Peter
Shor\cite{shor_c} and Andrew Steane\cite{steane_c}) are described.

Fault-tolerant processing of quantum information is the subject of \S4.  The
central challenge is to construct a universal set of quantum gates that can act
on the encoded data blocks without introducing an excessive number of errors.
Some schemes for universal computation (due to Peter Shor\cite{shor_c} and
Daniel Gottesman\cite{gottesman_b}) are outlined.

Once the elementary gates of our quantum computer are sufficiently reliable,
we can perform fault-tolerant quantum gates on encoded information, along with
fault-tolerant error recovery, to improve the reliability of the device.  But
for any fixed quantum code, or even for most infinite classes that contain
codes of arbitrarily large block size, these procedures will eventually fail if
we attempt a very long computation.  However, it is shown in \S5 that there is
a special class of codes ({\it concatenated codes}) which enable us to perform
longer and longer quantum computations reliably, as we increase the block size
at a modest rate.\cite{knill_a}$^{\!-\,}$\cite{zalka} Invoking concatenated
codes we can establish an {\it accuracy threshold} for quantum computation;
once our hardware meets a specified standard of accuracy, quantum
error-correcting codes and fault-tolerant procedures enable us to perform
arbitrarily long quantum computations with arbitrarily high reliability.  This
result is roughly analogous to Von Neumann's conclusion regarding classical
fault-tolerance, while the hierarchical structure of concatenated coding is
reminiscent of the G\'acs construction.  We outline an estimate of the accuracy
threshold, given assumptions about the errors that are enumerated in \S6.

With the development of fault tolerant methods, we now know that it is possible
in principle for the operator of a quantum computer to actively intervene to
stabilize the device against errors in a noisy (but not {\it too} noisy)
environment.  In the long term, though, fault tolerance might be achieved in
practical quantum computers by a rather different route---with intrinsically
fault-tolerant hardware.  Such hardware, designed to be impervious to {\it
localized} influences, could be operated relatively carelessly, yet could still
store and process quantum information robustly.  The topic of \S7 is a scheme
for fault-tolerant hardware envisioned by Alexei Kitaev,\cite{kitaev_c} in
which the quantum gates exploit nonabelian Aharonov-Bohm interactions among
distantly separated quasiparticles in a suitably constructed two-dimensional
spin system.  Though the laboratory implementation of Kitaev's idea may be far
in the future, his work offers a new slant on quantum fault tolerance that
shuns the analysis of abstract quantum circuits,  in favor of new physics
principles that might be exploited in the reliable processing of quantum
information.

The claims made in this chapter about the potential for the fault-tolerant
manipulation of complex quantum states may seem grandiose from the perspective
of present-day technology.  Surely, we have far to go before devices are
constructed that can, say, exploit the accuracy threshold for quantum
computation.  Nevertheless, I feel strongly that recent work relating to
quantum error correction will have an enduring legacy.
Theoretical quantum computation has developed at a spectacular pace over the
past three years.  If, as appears to be the case, the quantum classification of
computational complexity differs from the classical classification, then no
conceivable classical computer can accurately predict the behavior of even a
modest number of qubits (of order 100). Perhaps, then, relatively small quantum
systems will have far greater potential than we now suspect to surprise,
baffle, and delight us. Yet this potential could never be realized were we
unable to protect such systems from the destructive effects of noise and
decoherence.  Thus the discovery of fault-tolerant methods for quantum error
recovery and quantum computation has exceptionally deep implications, both for
the future of experimental physics and for the future of technology.  The
theoretical advances have illuminated the path toward a future in which
intricate quantum systems may be persuaded to do our bidding.

\section{Quantum error correction: the 7-qubit code}

To see how quantum error correction is possible, it is very instructive to
study a particular code. A simple and important example of a quantum
error-correcting code is the 7-qubit code devised by Andrew
Steane.~\cite{steane_a,steane_b}
This code enables us to store one qubit of quantum information (an arbitrary
state in a two-dimensional Hilbert space) using altogether 7-qubits (by
embedding the two-dimensional Hilbert space in a space of dimension $2^7$).
Steane's code is actually closely related to a familiar classical
error-correcting code, the [7,4,3] Hamming code.~\cite{macwilliams}
To understand why Steane's code works, it is important to first understand the
classical Hamming code.

The Hamming code uses a block of 7 bits to encode 4 bits of classical
information; that is, there are $16=2^4$ strings of length 7 that are the valid
codewords.  The codewords can be characterized by a parity check matrix 
\begin{equation}
\label{ham_matrix}
H=\pmatrix{0&0&0&1&1&1&1\cr
	0&1&1&0&0&1&1\cr
	1&0&1&0&1&0&1\cr} \ .
\end{equation}
Each valid codeword is a 7-bit string $v_{\rm code}$ that satisfies
\begin{equation}
\sum_k H_{jk}\left(v_{{\rm code}}\right)_k=0 ~({\rm mod ~ 2}) \ ;
\end{equation}
that is, the matrix $H$ annihilates each codeword in mod 2 arithmetic.  Since
$Z_2=\{0,1\}$ is a (finite) field, familiar results of linear algebra apply
here.   $H$ has three linearly independent rows and its kernel is spanned by
four linearly independent column vectors.  The 16 valid codewords are obtained
by taking all possible linear combinations of these four strings, with
coefficients chosen from $\{0,1\}$.

Now suppose that $v_{\rm code}$ is an (unknown) valid codeword, and that a
single (unknown) error occurs: one of the seven bits flips.  We are assigned
the task of determining which bit flipped, so that the error can be corrected.
This trick can be performed by applying the parity check matrix to the string.
Let $e_i$ denote the string with a one in the $i$th place, and zeros elsewhere.
 Then when the $i$th bit flips, $v_{\rm code}$ becomes $v_{\rm code}+e_i$.  If
we apply $H$ to this string we obtain
\begin{equation}
H\left(v_{\rm code}+ e_i\right)= H e_i
\end{equation}
(because $H$ annihilates $v_{\rm code}$), which is just the $i$th column of the
matrix $H$.  Since all of the columns of $H$ are distinct, we can infer $i$; we
have learned where the error occurred, and we can correct the error by flipping
the $i$th bit back.  Thus, we can recover the encoded data unambiguously if
only one bit flips; but if two or more different bits flip, the encoded data
will be damaged.  It is noteworthy that the quantity $H e_i$ reveals the
location of the error without telling us anything about $v_{\rm code}$; that
is, without revealing the encoded information.

Steane's code generalizes this sort of classical error-correcting code to a
{\it quantum} error-correcting code.  The code uses a 7-qubit ``block'' to
encode one qubit of quantum information, that is, we can encode an arbitrary
state in a two-dimensional Hilbert space spanned by two states: the ``logical
zero'' $|0\rangle_{\rm code}$ and the ``logical one'' $|1\rangle_{\rm code}$.
The code is designed to enable us to recover from an arbitrary error occurring
in any of the 7 qubits in the block.

What do we mean by an arbitrary error?  The qubit in question might undergo a
random {\it unitary} transformation, or it might {\it decohere} by becoming
entangled with states of the environment.  Suppose that, if no error occurs,
the qubit ought be in the state $a|0\rangle +b|1\rangle$.  (Of course, this
particular qubit might be entangled with others, so the coefficients $a$ and
$b$ need not be complex numbers; they can be states that are orthogonal to both
$|0\rangle$ and $|1\rangle$, which we assume are unaffected by the error.)  Now
if the qubit is afflicted by an arbitrary error, the resulting state can be
expanded in the form:
\begin{eqnarray}
\label{superop}
a|0\rangle + b|1\rangle\longrightarrow &\left(a|0\rangle +
b|1\rangle\right)&\otimes \quad |A_{\rm no~error}~~~\rangle_{\rm
env}\nonumber\\
+&\left(a|1\rangle + b|0\rangle\right)&\otimes \quad |A_{\rm
bit-flip}~~~~\rangle_{\rm env}\nonumber\\
+&\left(a|0\rangle - b|1\rangle\right)&\otimes \quad |A_{\rm
phase-flip}~\rangle_{\rm env}\nonumber\\
+&\left(a|1\rangle - b|0\rangle\right)&\otimes \quad 
|A_{\rm both~errors}\rangle_{\rm env} \ ,\nonumber\\
\end{eqnarray}
where each $|A\rangle_{\rm env}$ denotes a state of the environment.  We are
making no particular assumption about the orthogonality or normalization of the
$|A\rangle_{\rm env}$ states,\footnote{Though, of course, the combined
evolution of qubit plus environment is required to be unitary.} so
Eq.~(\ref{superop}) entails no loss of generality.  We conclude that the
evolution of the qubit can be expressed as a linear combination of four
possibilities:  (1) no error occurs, (2) the bit flip $|0\rangle
\leftrightarrow |1\rangle$ occurs, (3) the relative phase of $|0\rangle$ and
$|1\rangle$ flips, (4) both a bit flip and a phase flip occur.

Now it is clear how a quantum error-correcting code should
work.\cite{steane_b,knill_b} By making a suitable measurement, we wish to
diagnose
which of these four possibilities actually occurred.  Of course, in general,
the state of the qubit will be a linear combination of these four states, but
the measurement should project the state onto the basis used in
Eq.~(\ref{superop}).  We can then proceed to correct the error by applying one
of
the four unitary transformations:
\begin{equation}
\label{xyz_define}
(1)~ {\bf 1} \ , \quad (2) ~ X\equiv\pmatrix{0&1\cr 1&0\cr} \ , \quad (3) ~
Z\equiv\pmatrix{1&0\cr 0& -1\cr} \ , \quad (4)~  Y\equiv X\cdot Z \ ,
\end{equation}
(and the measurement outcome will tell us which one to apply).
By applying this transformation, we restore the qubit to its intended value,
and completely disentangle the quantum state of the qubit from the state of the
environment. It is essential, though, that in diagnosing the error, we learn
nothing about the encoded quantum information, for to find out anything about
the coefficients $a$ and $b$ in Eq.~(\ref{superop}) would necessarily destroy
the
coherence of the qubit.

\begin{figure}
\centering
\begin{picture}(360,80)

\put(0,36){\makebox(20,12){$x$}}
\put(20,40){\line(1,0){40}}
\put(40,40){\circle{8}}
\put(40,44){\line(0,-1){8}}
\put(70,36){\makebox(20,12){$x\oplus 1$}}

\put(120,46){\makebox(20,12){$x$}}
\put(120,26){\makebox(20,12){$y$}}
\put(140,50){\line(1,0){40}}
\put(140,30){\line(1,0){40}}

\put(160,50){\circle*{4}}
\put(160,50){\line(0,-1){24}}
\put(160,30){\circle{8}}

\put(190,26){\makebox(20,12){$x\oplus y$}}
\put(190,46){\makebox(20,12){$x$}}

\put(240,56){\makebox(20,12){$x$}}
\put(240,36){\makebox(20,12){$y$}}
\put(240,16){\makebox(20,12){$x$}}
\put(260,60){\line(1,0){40}}
\put(260,40){\line(1,0){40}}
\put(260,20){\line(1,0){40}}

\put(280,60){\circle*{4}}
\put(280,40){\circle*{4}}
\put(280,20){\circle{8}}
\put(280,60){\line(0,-1){44}}

\put(310,56){\makebox(20,12){$x$}}
\put(310,36){\makebox(20,12){$y$}}
\put(310,16){\makebox(20,12){$z\oplus xy$}}

\end{picture}
\caption{Diagrammatic notation for the NOT gate, the XOR (controlled-NOT) gate,
and the Toffoli (controlled-controlled-NOT) gate.}
\label{fig_gates}
\end{figure}
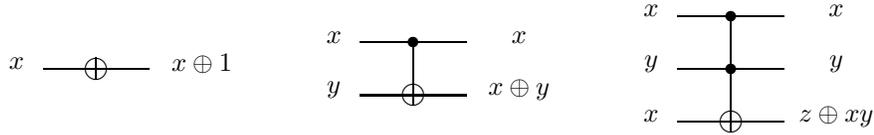

\begin{figure}
\centering
\begin{picture}(320,115)

\put(0,30){\makebox(0,0){$\ket{0}$}}
\put(70,25){\makebox(20,12){Measure}}

\put(10,110){\line(1,0){270}}
\put(10,100){\line(1,0){270}}
\put(10,90){\line(1,0){270}}
\put(10,80){\line(1,0){270}}
\put(10,70){\line(1,0){270}}
\put(10,60){\line(1,0){270}}
\put(10,50){\line(1,0){270}}

\put(10,30){\line(1,0){50}}

\put(20,80){\circle*{4}}
\put(20,80){\line(0,-1){54}}
\put(20,30){\circle{8}}

\put(30,70){\circle*{4}}
\put(30,70){\line(0,-1){44}}
\put(30,30){\circle{8}}

\put(40,60){\circle*{4}}
\put(40,60){\line(0,-1){34}}
\put(40,30){\circle{8}}

\put(50,50){\circle*{4}}
\put(50,50){\line(0,-1){24}}
\put(50,30){\circle{8}}

\put(110,20){\makebox(0,0){$\ket{0}$}}
\put(180,15){\makebox(20,12){Measure}}

\put(120,20){\line(1,0){50}}

\put(130,100){\circle*{4}}
\put(130,100){\line(0,-1){84}}
\put(130,20){\circle{8}}

\put(140,90){\circle*{4}}
\put(140,90){\line(0,-1){74}}
\put(140,20){\circle{8}}

\put(150,60){\circle*{4}}
\put(150,60){\line(0,-1){44}}
\put(150,20){\circle{8}}

\put(160,50){\circle*{4}}
\put(160,50){\line(0,-1){34}}
\put(160,20){\circle{8}}

\put(220,10){\makebox(0,0){$\ket{0}$}}
\put(290,5){\makebox(20,12){Measure}}

\put(230,10){\line(1,0){50}}

\put(240,110){\circle*{4}}
\put(240,110){\line(0,-1){104}}
\put(240,10){\circle{8}}

\put(250,90){\circle*{4}}
\put(250,90){\line(0,-1){84}}
\put(250,10){\circle{8}}

\put(260,70){\circle*{4}}
\put(260,70){\line(0,-1){64}}
\put(260,10){\circle{8}}

\put(270,50){\circle*{4}}
\put(270,50){\line(0,-1){44}}
\put(270,10){\circle{8}}

\end{picture}
\caption{Computation of the bit-flip syndrome for Steane's 7-qubit code.
Repeating the  computation in the rotated basis diagnoses the phase-flip
errors.  To make the procedure fault tolerant, each ancilla qubit must be
replaced by four qubits in a suitable state.}
\label{fig_syndrome}
\end{figure}
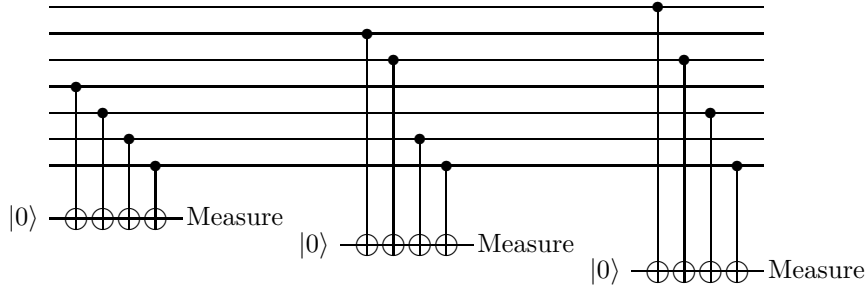

If we use Steane's code, a measurement meeting these criteria is possible.  The
logical zero is the equally weighted superposition of all of the even weight
codewords of the Hamming code (those with an even number of 1's), 
\begin{eqnarray}
\label{zero}
 |0\rangle_{\rm code}&=&
{1\over\sqrt{8}} \left(\sum_{{\rm even}~v\atop \in ~{\rm
Hamming}}|v\rangle\right)\nonumber\\
&={1\over \sqrt{8}} & \Big(  |0000000\rangle + |0001111\rangle +|0110011\rangle
+|0111100\rangle\nonumber\\
 && +  |1010101\rangle  +|1011010\rangle +|1100110\rangle +|1101001\rangle\Big)
\ ,\nonumber\\
\end{eqnarray}
and the logical 1 is the equally weighted superposition of all of the odd
weight codewords of the Hamming code (those with an odd number of 1's),
\begin{eqnarray}
\label{one}
|1\rangle_{\rm code}&=& {1\over\sqrt{8}}\left(\sum_{{\rm odd}~v\atop \in ~{\rm
Hamming}}|v\rangle \right)\nonumber\\
&={1\over \sqrt{8}} &\Big(   |1111111\rangle  + |1110000\rangle
+|1001100\rangle
+|1000011\rangle\nonumber\\
&& + |0101010\rangle  +|0100101\rangle +|0011001\rangle +|0010110\rangle\Big) \
.\nonumber \\ 
\end{eqnarray}
Since all of the states appearing in Eq.~(\ref{zero}) and Eq.~(\ref{one}) are
Hamming codewords, it is easy to detect a single bit flip in the block by doing
a simple quantum computation, as illustrated in Fig.~\ref{fig_syndrome} (using
notation defined in Fig~\ref{fig_gates}). We
augment the block of 7 qubits with 3 ancilla bits,\footnote{To make the
procedure fault-tolerant, we will need to increase the number of ancilla bits
as discussed in \S3.} and perform the unitary operation:
\begin{equation}
|v\rangle\otimes|0\rangle_{\rm anc}\longrightarrow
|v\rangle\otimes|Hv\rangle_{\rm anc} \ ,
\end{equation}
where $H$ is the Hamming parity check matrix, and $|\cdot\rangle_{\rm anc}$
denotes the state of the three ancilla bits. If we assume that only a single
one of the 7 qubits in the block is in error, measuring the ancilla projects
that qubit onto either a state with a bit flip or a state with no flip (rather
than any nontrivial superposition of the two). If the bit does flip, the
measurement outcome diagnoses which bit was affected, without revealing
anything about the quantum information encoded in the block.  

But to perform quantum error correction, we will need to diagnose phase errors
as well as bit flip errors.  To accomplish this, we observe (following
Steane\cite{steane_a,steane_b}) that we can change the basis for each qubit by
applying the Hadamard
rotation
\begin{equation}
R= {1\over\sqrt{2}}\pmatrix{1&1\cr 1&-1\cr} \ .
\end{equation}
Then phase errors in the $|0\rangle$, $|1\rangle$ basis become bit flip errors
in the rotated basis
\begin{equation}
|\tilde 0\rangle \equiv {1\over\sqrt{2}}\left(|0\rangle +|1\rangle\right) \
,\quad 
|\tilde 1\rangle \equiv {1\over\sqrt{2}}\left(|0\rangle -|1\rangle\right) \ .
\end{equation}
It will therefore be sufficient if our code is able to diagnose bit flip errors
in this rotated basis. But if we apply the Hadamard rotation to each of the 7
qubits, then Steane's logical 0 and logical 1 become in the rotated basis
\begin{eqnarray}
\label{tildezero}
|\tilde 0\rangle_{\rm code}=&{1\over 4}\left(\sum_{v \in \atop {\rm
Hamming}}|v\rangle\right)
={1\over \sqrt{2}}\left(|0\rangle_{\rm code} + |1\rangle_{\rm code}\right) \
,\nonumber\\
|\tilde 1\rangle_{\rm code}=&{1\over 4}\left(\sum_{v \in \atop {\rm
Hamming}}(-1)^{wt(v)}|v\rangle \right)
={1\over \sqrt{2}}\left(|0\rangle_{\rm code} - |1\rangle_{\rm code}\right)\
\nonumber\\
\end{eqnarray}
(where $wt(v)$ denotes the weight of $v$). The key point is that $|\tilde
0\rangle_{\rm code}$ and $|\tilde 1\rangle_{\rm code}$, like $|0\rangle_{\rm
code}$ and $|1\rangle_{\rm code}$, are superpositions of Hamming codewords.
Hence, in the rotated basis, as in the original basis, we can perform the
Hamming parity check to diagnose bit flips, which are phase flips in the
original basis.  Assuming that only one qubit is in error, performing the
parity check in both bases completely diagnoses the error, and enables us to
correct it.

In the above description of the error correction scheme, I assumed that the
error affected only one of the qubits in the block.  Clearly, this assumption
as stated is not realistic; all of the qubits will typically become entangled
with the environment to some degree.   However, as we have seen, the procedure
for determining the error syndrome will typically project each qubit onto a
state in which no error has occurred.  For each qubit, there is a non-zero
probability of an error, assumed small, which we'll call $\epsilon$.  Now we
will make a very important assumption -- that the errors acting on different
qubits in the same block are completely uncorrelated with one another.  Under
this assumption, the probability of two errors is of order $\epsilon^2$, and so
is much smaller than the probability of a single error if $\epsilon$ is small
enough.  So, to order $\epsilon$ accuracy, we can safely confine our attention
to the case where at most one qubit per block is in error. (In fact, to reach
this conclusion, we do not really require that errors acting on different
qubits be {\it completely} uncorrelated.  If all qubits are exposed to the same
weak magnetic field, so that each has a probability $\epsilon$ of flipping
over, that would be okay because the probability that two spins flip over is
order $\epsilon^2$.  What would cause trouble is a process occuring with
probability of order $\epsilon$ that flips two spins at once.)

But in the (unlikely) event of two errors occurring in the same block of the
code, our recovery procedure will typically fail.  If two bits flip in the same
block, then the Hamming parity check will misdiagnose the error.  Recovery will
restore the quantum state to the code subspace, but the {\it encoded}
information in the block will undergo the bit flip
\begin{equation}
|0\rangle_{\rm code} \rightarrow |1\rangle_{\rm code} \ ,
\quad |1\rangle_{\rm code} \rightarrow |0\rangle_{\rm code} \ .
\end{equation}
Similarly, if there are two phase errors in the same block, these are two bit
flip errors in the rotated basis, so that after recovery the block will have
undergone a bit flip in the rotated basis, or in the original basis the phase
flip
\begin{equation}
|0\rangle_{\rm code} \rightarrow |0\rangle_{\rm code} \ ,\quad |1\rangle_{\rm
code} \rightarrow -|1\rangle_{\rm code} \ .
\end{equation}
(If one qubit in the block has a phase error, and another one has a bit flip
error, then recovery will be successful.)

Thus we have seen that Steane's code can enhance the reliability of stored
quantum information.  Suppose that we want to store one qubit in an unknown
pure state $|\psi\rangle$.  Due to imperfections in our storage device, the
state $\rho_{\rm out}$ that we recover will have suffered a loss of fidelity:
\begin{equation}
F\equiv \langle\psi|\rho_{\rm out}|\psi\rangle=1-\epsilon \ .
\end{equation}
But if we store the qubit using Steane's 7-qubit block code, if each of the
7-qubits is maintained with fidelity $F=1-\epsilon$, if the errors on the
qubits are uncorrelated, and if we can perform error recovery, encoding, and
decoding  flawlessly (more on this below), then the encoded information can be
maintained with an improved fidelity $F=1- O\left(\epsilon^2\right)$.  

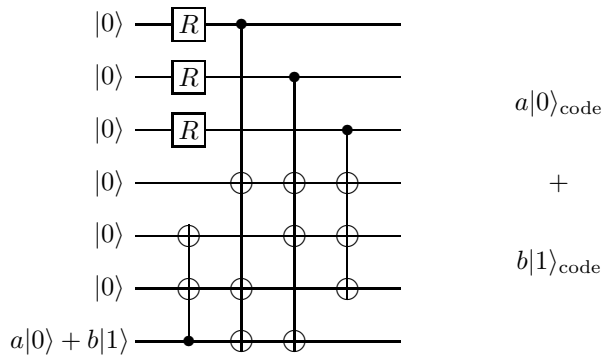
\begin{figure}
\centering
\begin{picture}(260,140)

\put(25,10){\makebox(0,0){$a \ket{0}+ b \ket{1}$}}
\put(40,30){\makebox(0,0){$\ket{0}$}}
\put(40,50){\makebox(0,0){$\ket{0}$}}
\put(40,70){\makebox(0,0){$\ket{0}$}}
\put(40,90){\makebox(0,0){$\ket{0}$}}
\put(40,110){\makebox(0,0){$\ket{0}$}}
\put(40,130){\makebox(0,0){$\ket{0}$}}

\put(50,10){\line(1,0){100}}
\put(50,30){\line(1,0){100}}
\put(50,50){\line(1,0){100}}
\put(50,70){\line(1,0){100}}
\put(50,130){\line(1,0){14}}
\put(50,110){\line(1,0){14}}
\put(50,90){\line(1,0){14}}

\put(70,10){\circle*{4}}
\put(70,10){\line(0,1){44}}
\put(70,30){\circle{8}}
\put(70,50){\circle{8}}

\put(64,124){\framebox(12,12){$R$}}
\put(64,104){\framebox(12,12){$R$}}
\put(64,84){\framebox(12,12){$R$}}

\put(76,130){\line(1,0){74}}
\put(76,110){\line(1,0){74}}
\put(76,90){\line(1,0){74}}

\put(90,130){\circle*{4}}
\put(90,130){\line(0,-1){124}}
\put(90,10){\circle{8}}
\put(90,30){\circle{8}}
\put(90,70){\circle{8}}

\put(110,110){\circle*{4}}
\put(110,110){\line(0,-1){104}}
\put(110,10){\circle{8}}
\put(110,50){\circle{8}}
\put(110,70){\circle{8}}

\put(130,90){\circle*{4}}
\put(130,90){\line(0,-1){64}}
\put(130,30){\circle{8}}
\put(130,50){\circle{8}}
\put(130,70){\circle{8}}

%\put(180,70){\makebox(0,0){\scalebox{5}{\Huge $\}$}}}

\put(210,100){\makebox(0,0){$a|0\rangle_{\rm code}$}}
\put(210,70){\makebox(0,0){+}}
\put(210,40){\makebox(0,0){$b|1\rangle_{\rm code}$}}

\end{picture}
\caption{An encoding circuit for Steane's 7-qubit code.}
\label{fig_encode}
\end{figure}

A qubit in an unknown state can be encoded using the circuit shown in
Fig.~\ref{fig_encode}.  It is easiest to understand how the encoder works by
using an alternative expression for the Hamming parity check matrix,
\begin{equation}
\label{another_ham}
H=\pmatrix{1&0&0&1&0&1&1\cr
	0&1&0&1&1&0&1\cr
	0&0&1&1&1&1&0\cr} \ .
\end{equation}
(This form of $H$ is obtained from the form in Eq.~(\ref{ham_matrix}) by
permuting the columns, which is just a relabeling of the bits in the block.)
The even subcode of the Hamming code is actually the space spanned by the rows
of $H$; so we see that (in this representation of $H$) the first three bits of
the string completely characterize the data represented in the subcode.  The
remaining four bits are the parity bits that provide the redundancy needed to
protect against errors.  When encoding the unknown state $a |0\rangle+
b|1\rangle$, the encoder first uses two XOR's to prepare the state $a
|0000000\rangle+ b|0000111\rangle$, a superposition of even and odd Hamming
codewords. The rest of the circuit adds $|0\rangle_{\rm code}$ to this state:
the Hadamard ($R$) rotations prepare an equally weighted superposition of all
eight possible values for the first three bits in the block, and the remaining
XOR gates switch on the parity bits dictated by $H$. 

We will also want to be able to measure the encoded qubit, say by projecting
onto the orthogonal basis $\{|0\rangle_{\rm code}, |1\rangle_{\rm code}\}$. If
we don't mind destroying the encoded block when we make the measurement, then
it is sufficient to measure each of the seven qubits in the block by projecting
onto the basis $\{|0\rangle,|1\rangle\}$; we then perform classical error
correction on the measurement outcomes to obtain a Hamming codeword.  The
parity of that codeword is the value of the logical qubit.  (The classical
error correction step provides protection against measurement errors. For
example, if the block is in the state $|0\rangle_{\rm code}$, then two
independent errors would have to occur in the measurement of the elementary
qubits for the measurement of the logical qubit to yield the incorrect value
$|1\rangle_{\rm code}$.)

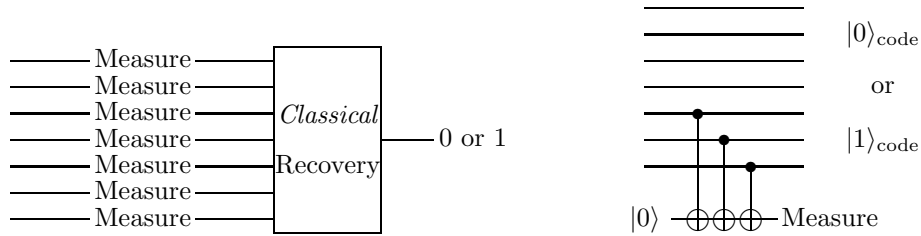
\begin{figure}
\centering
\begin{picture}(360,115)

\put(0,70){\line(1,0){30}}
\put(0,60){\line(1,0){30}}
\put(0,50){\line(1,0){30}}
\put(0,40){\line(1,0){30}}
\put(0,30){\line(1,0){30}}
\put(0,20){\line(1,0){30}}
\put(0,10){\line(1,0){30}}

\put(40,65){\makebox(20,12){Measure}}
\put(40,55){\makebox(20,12){Measure}}
\put(40,45){\makebox(20,12){Measure}}
\put(40,35){\makebox(20,12){Measure}}
\put(40,25){\makebox(20,12){Measure}}
\put(40,15){\makebox(20,12){Measure}}
\put(40,5){\makebox(20,12){Measure}}

\put(70,70){\line(1,0){30}}
\put(70,60){\line(1,0){30}}
\put(70,50){\line(1,0){30}}
\put(70,40){\line(1,0){30}}
\put(70,30){\line(1,0){30}}
\put(70,20){\line(1,0){30}}
\put(70,10){\line(1,0){30}}

\put(100,5){\framebox(40,70){}}
\put(120,50){\makebox(0,0){\it Classical}}
\put(120,30){\makebox(0,0){Recovery}}

\put(140,40){\line(1,0){20}}
\put(165,35){\makebox(20,12){0 or 1}}

\put(240,90){\line(1,0){60}}
\put(240,80){\line(1,0){60}}
\put(240,70){\line(1,0){60}}
\put(240,60){\line(1,0){60}}
\put(240,50){\line(1,0){60}}
\put(240,40){\line(1,0){60}}
\put(240,30){\line(1,0){60}}

\put(250,10){\line(1,0){40}}

\put(240,10){\makebox(0,0){$\ket{0}$}}
\put(300,5){\makebox(20,12){Measure}}

\put(260,50){\circle*{4}}
\put(260,50){\line(0,-1){44}}
\put(260,10){\circle{8}}

\put(270,40){\circle*{4}}
\put(270,40){\line(0,-1){34}}
\put(270,10){\circle{8}}

\put(280,30){\circle*{4}}
\put(280,30){\line(0,-1){24}}
\put(280,10){\circle{8}}

\put(330,80){\makebox(0,0){$|0\rangle_{\rm code}$}}
\put(330,60){\makebox(0,0){or}}
\put(330,40){\makebox(0,0){$|1\rangle_{\rm code}$}}

%\put(285,60){\makebox(0,0){\Huge $\}$}}

\end{picture}
\caption{Destructive and nondestructive measurement of the logical qubit.}
\label{fig_measure}
\end{figure}

In applications to quantum computation, we will need to perform a measurement
that projects onto $\{|0\rangle_{\rm code}, |1\rangle_{\rm code}\}$ without
destroying the block.  This task is accomplished by copying the parity of the
block onto an ancilla qubit, and then measuring the ancilla. A circuit that
performs a nondestructive measurement of the code block (in the case where the
parity check matrix is as in Eq.~(\ref{another_ham})) is shown in
Fig.~\ref{fig_measure}.  The measurement is nondestructive in the sense that it
preserves the code subspace; it does, of course ``destroy'' a coherent
superposition
$ a|0\rangle_{\rm code} + b|1\rangle_{\rm code}$ by collapsing the state to
either
$|0\rangle_{\rm code}$ (with probability $|a|^2$) or $|1\rangle_{\rm code}$
(with probability $|b|^2$).

Steane's 7-qubit code can recover from only a single error in the code block,
but better codes can be
constructed\cite{steane_b}$^{\!,\,}$\cite{calderbank_a}$^{\!-\,}$\cite{calderbank_c} that can protect the information from up to
$t$ errors within a single block, so that the encoded information can be
maintained with a fidelity $F=1 - O\left(\epsilon^{t+1}\right)$.
The current status of quantum coding theory is reviewed by Steane in this
volume.

The key conceptual insight that makes quantum error correction possible is that
we
can {\it fight entanglement with entanglement}.
Entanglement can be our enemy, since entanglement of our device with the
environment can conceal quantum information from us, and so cause errors.  But
entanglement can also be our friend---we can encode the information that we
want to protect in entanglement, that is, in correlations involving a large
number of qubits.  This information, then, cannot be accessed if we measure
just a few qubits.  By the same token, the information cannot be {\it
damaged} if the environment interacts with just a few qubits.

Furthermore, we
have learned that, although the quantum computer is in a sense an analog
device, we can {\it digitalize} the errors that it makes.  We deal with small
errors by making appropriate measurements that project the state of our quantum
computer onto either a state where no error has occurred, or a state with a
large error, which can then be corrected with familiar methods.  And we have
seen that it is possible to measure the errors without measuring the data---we
can acquire information about the precise nature of the error without acquiring
any information about the quantum information encoded in our device (which
would result in decoherence and failure of our computation).

All quantum error correcting codes make use of the same fundamental strategy:
a small subspace of the Hilbert space of
our device is designated as the {\it code subspace}.  This space is carefully
chosen so that all of the errors that we want to correct move the code space to
mutually orthogonal {\it error subspaces}.  We can make a measurement after our
system has interacted with the environment that tells us in which of these
mutually orthogonal subspaces the system resides, and hence infer exactly what
type of error occurred.  The error can then be repaired by applying an
appropriate unitary transformation.

\section{Fault-tolerant recovery}
\label{sec_recovery}

In our discussion so far, we have assumed that we can encode quantum
information and perform recovery from errors without making any mistakes. But,
of course, error recovery will not be flawless.  Recovery is itself a
quantum computation that will be prone to error.  If the probability of error
for each bit in our code block is $\epsilon$, then it is reasonable to suppose
that each quantum gate that we employ in the recovery procedure has a
probability of order $\epsilon$ of introducing an error (or that ``storage
errors'' occur with probability of order $\epsilon$ during recovery).   If our
recovery procedure is carelessly designed, then the probability that the
procedure fails ({\it e.g.}, because two errors occur in the same block) may be
of order $\epsilon$.  Then we have derived no benefit from using a quantum
error-correcting code; in fact, the probability of error per data qubit is even
higher than without any coding.  So we are obligated to consider systematically
all the possible ways that recovery might fail with a probability of order
$\epsilon$, and to ensure that they are all eliminated.  Only then is our
procedure {\it fault tolerant}, and only then is coding guaranteed to pay off
once $\epsilon$ is small enough.

\subsection{The Back Action Problem}

One serious concern is propagation of error.  If an error occurs in one qubit,
and then we apply a gate in which that qubit interacts with another, the error
is likely to spread to the second qubit.  We need to be careful to contain the
infection, or at least we must strive to prevent two errors from appearing in a
single block.

\begin{figure}
\centering
\begin{picture}(210,85)

\put(10,30){\line(1,0){19}}
\put(29,24){\framebox(12,12){$R$}}
\put(41,30){\line(1,0){38}}

\put(10,70){\line(1,0){19}}
\put(29,64){\framebox(12,12){$R$}}
\put(41,70){\line(1,0){38}}

\put(79,24){\framebox(12,12){$R$}}
\put(91,30){\line(1,0){19}}

\put(79,64){\framebox(12,12){$R$}}
\put(91,70){\line(1,0){19}}

\put(60,70){\circle*{6}}
\put(60,70){\line(0,-1){45}}
\put(60,30){\circle{10}}

\put(120,46){\makebox(20,12){\Large\bf =}}

\put(150,30){\line(1,0){60}}
\put(150,70){\line(1,0){60}}
\put(180,30){\circle*{6}}
\put(180,30){\line(0,1){45}}
\put(180,70){\circle{10}}

\end{picture}
\caption{A useful identity. The source and the target of an XOR gate are
interchanged if we perform a change of basis with Hadamard rotations.}
\label{identity}
\end{figure}
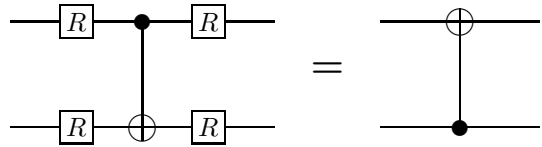

In performing error recovery, we repeatedly use the two-qubit XOR gate.  This
gate can propagate errors in two different ways.  First, it is obvious that if
a bit flip error occurs in one qubit, and that qubit is then used as the source
qubit of an XOR gate, then the bit flip will propagate ``forward'' to the
target qubit.  The second type of error propagation is more subtle, and can be
understood using the identity represented in Fig.~\ref{identity} --- if we
perform a rotation of basis with a Hadamard gate on both qubits, then the
source and the target of the XOR gate are interchanged.  Since we recall that
this change of basis also interchanges a bit flip error with a phase error, we
infer that if a phase error occurs in one qubit, and that qubit is then used as
the {\it target} qubit of an XOR gate, then the error will propagate
``backward'' to the source qubit.

We can now see that the circuit shown in Fig.~\ref{fig_syndrome} is {\it not}
fault tolerant.   The trouble is that a single ancilla qubit is used as a
target for four successive XOR gates.  If just a single phase error occurs in
the ancilla qubit at some stage, that one error can feed back to two or more of
the qubits in the data block.  The result is that a block phase error may occur
with a probability of order $\epsilon$, which is not acceptable.

\begin{figure}
\centering
\begin{picture}(320,105)

\put(60,0){\makebox(20,12){\bf Bad!}}
\put(0,46){\makebox(20,12){Ancilla}}
\put(0,81){\makebox(20,12){Data}}

\put(30,50){\line(1,0){100}}

\put(30,70){\line(1,0){100}}
\put(30,80){\line(1,0){100}}
\put(30,90){\line(1,0){100}}
\put(30,100){\line(1,0){100}}

\put(50,100){\circle*{4}}
\put(50,100){\line(0,-1){54}}
\put(50,50){\circle{8}}

\put(70,90){\circle*{4}}
\put(70,90){\line(0,-1){44}}
\put(70,50){\circle{8}}

\put(90,80){\circle*{4}}
\put(90,80){\line(0,-1){34}}
\put(90,50){\circle{8}}

\put(110,70){\circle*{4}}
\put(110,70){\line(0,-1){24}}
\put(110,50){\circle{8}}

\put(250,0){\makebox(20,12){\bf Good!}}
\put(190,31){\makebox(20,12){Ancilla}}
\put(190,81){\makebox(20,12){Data}}

\put(220,20){\line(1,0){100}}
\put(220,30){\line(1,0){100}}
\put(220,40){\line(1,0){100}}
\put(220,50){\line(1,0){100}}

\put(220,70){\line(1,0){100}}
\put(220,80){\line(1,0){100}}
\put(220,90){\line(1,0){100}}
\put(220,100){\line(1,0){100}}

\put(240,100){\circle*{4}}
\put(240,100){\line(0,-1){54}}
\put(240,50){\circle{8}}

\put(260,90){\circle*{4}}
\put(260,90){\line(0,-1){54}}
\put(260,40){\circle{8}}

\put(280,80){\circle*{4}}
\put(280,80){\line(0,-1){54}}
\put(280,30){\circle{8}}

\put(300,70){\circle*{4}}
\put(300,70){\line(0,-1){54}}
\put(300,20){\circle{8}}

\end{picture}
\caption{Bad and good versions of syndrome measurement. The bad circuit uses
the same ancilla bit several
times; the good circuit uses each ancilla bit only once.}
\label{fig:good_bad}
\end{figure}
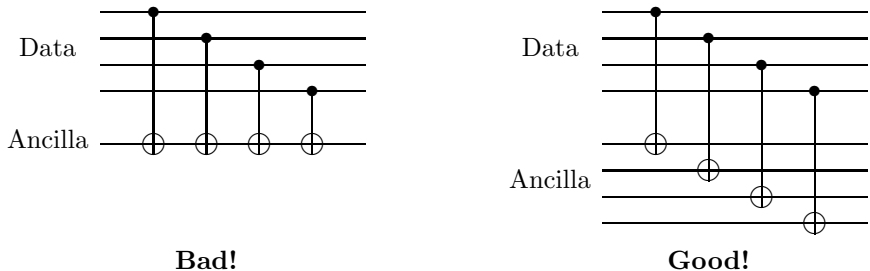

To reduce the failure probability to order $\epsilon^2$, we must modify the
recovery circuit so that each ancilla qubit couples to no more than one qubit
within the code block.  One way to do this is to expand the ancilla from one
bit to four, with each bit the target of a single XOR gate, as in
Fig.~\ref{fig:good_bad}.  We can then measure all four ancilla bits.  The bit
of the syndrome that we are seeking is the {\it parity} of the four measured
bits.  In effect, we have copied from the data block to the ancilla some
information about the error that occurred,  and we read that information when
we measure the ancilla.

But this procedure is still not adequate as it stands, because we have copied
{\it too much} information.  The circuit entangles the ancilla with the error
that has occured in the data, which is good, but it also entangles the ancilla
with the encoded data itself, which is bad.  The measurement of the ancilla
destroys the carefully prepared superposition of basis states in the
expressions Eqs.~(\ref{zero}) and ({\ref{one}) for $|0\rangle_{\rm code}$ and
$|1\rangle_{\rm code}$.  For example, suppose we are measuring the first bit of
the syndrome as in Fig.~\ref{fig_syndrome}, but with the ancilla expanded from
one bit to four.  In effect, then, we are measuring the last four bits of the
block.  If we obtain the measurement result, say, $|0000\rangle_{\rm anc}$,
then we have projected $|0\rangle_{\rm code}$ to $|0000000\rangle$ and
$|1\rangle_{\rm code}$ to $|1110000\rangle$; the codewords have lost all
protection against phase errors.

\subsection{Preparing the Ancilla}

We need to modify the recovery procedure further, preserving its good features
while eliminating its bad features.  We want to copy onto our ancilla the
information about the errors in the data block, without feeding multiple phase
errors into the data, and without destroying the coherence of the data. To meet
this goal, we must prepare an appropriate state of the ancilla before the error
syndrome computation begins.  This state is chosen so the outcome of the
ancilla measurement will reveal the information about the errors without
revealing anything about the state of the data.

One way to meet this criterion was suggested by Peter Shor; The {\it Shor
state} that he proposed is a state of four
ancilla bits that is an equally weighted superposition of all even weight
strings:
\begin{equation}
|{\rm Shor}\rangle_{\rm anc}={1\over\sqrt{8}}\sum_{{\rm even}~v}|v\rangle_{\rm
anc} \ .
\end{equation}
To compute each bit of the syndrome, we prepare the ancilla in a Shor state,
perform four XOR
gates (with appropriate qubits in the data block as the sources and the four
bits of the Shor state as the targets), and then measure the ancilla state.

If the syndrome bit we are computing is trivial, then the computation adds an
even weight string to the Shor state, which leaves it unchanged; if the
syndrome bit is nontrivial, the Shor state is transformed to the equally
weighted superposition of odd weight strings.  Thus, the parity of the
measurement result reveals the value of the syndrome bit, but no other
information about the state of the data block can be extracted from measurement
--- we {\it have} found a way to extract the syndrome without damaging the
codewords.  (The particular string of specified parity that we find in the
measurement is selected at random, and has nothing to do with the state of the
data block.)

There are altogether 6 syndrome bits (3 to diagnose bit-flip errors and 3 to
diagnose phase-flip errors), so the syndrome measurement uses 24 ancilla bits
prepared in 6 Shor states, and 24 XOR gates.

One way to obtain the phase-flip syndrome would be to first apply 7 parallel
$R$ gates to the data block to rotate the basis, then to apply the XOR gates as
in Fig.~\ref{fig_syndrome} (but with the ancilla expanded into a Shor state),
and finally to apply 7 $R$ gates to rotate the data back.  However, we can use
the identity represented in Fig.~\ref{identity} to improve this procedure.  By
reversing the direction of the XOR gates (that is, by using the ancilla as the
source and the data as the target), we can avoid applying the $R$ gates to the
data, and hence can reduce the likelihood of damaging the data with faulty
gates,\cite{zalka,steane_c} as shown in Fig.~\ref{fig:synd_circuits}.

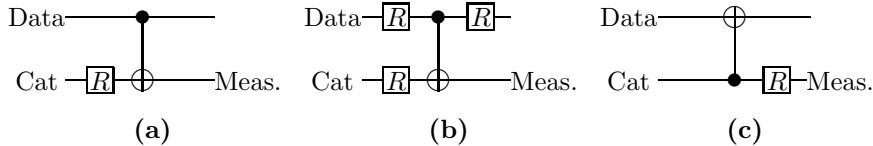
\begin{figure}
\centering
\begin{picture}(400,65)
\setlength{\unitlength}{.8pt}
\put(60,0){\makebox(10,12){\bf (a)}}
\put(200,0){\makebox(10,12){\bf (b)}}
\put(340,0){\makebox(10,12){\bf (c)}}

%% (a)
\put(0,24){\makebox(20,12){Cat}}
\put(0,54){\makebox(20,12){Data}}

\put(24,30){\line(1,0){10}}
\put(34,24){\framebox(12,12){$R$}}
\put(46,30){\line(1,0){48}}

\put(24,60){\line(1,0){70}}

\put(60,60){\circle*{6}}
\put(60,60){\line(0,-1){35}}
\put(60,30){\circle{10}}

\put(100,24){\makebox(20,12){Meas.}}

%%(b)
\put(140,24){\makebox(20,12){Cat}}
\put(140,54){\makebox(20,12){Data}}

\put(164,30){\line(1,0){10}}
\put(174,24){\framebox(12,12){$R$}}
\put(186,30){\line(1,0){48}}

\put(164,60){\line(1,0){10}}
\put(174,54){\framebox(12,12){$R$}}
\put(186,60){\line(1,0){28}}
\put(214,54){\framebox(12,12){$R$}}
\put(226,60){\line(1,0){8}}

\put(200,60){\circle*{6}}
\put(200,60){\line(0,-1){35}}
\put(200,30){\circle{10}}

\put(240,24){\makebox(20,12){Meas.}}

%%(c)
\put(280,24){\makebox(20,12){Cat}}
\put(280,54){\makebox(20,12){Data}}

\put(304,30){\line(1,0){50}}
\put(354,24){\framebox(12,12){$R$}}
\put(366,30){\line(1,0){8}}

\put(304,60){\line(1,0){72}}

\put(340,30){\circle*{6}}
\put(340,30){\line(0,1){35}}
\put(340,60){\circle{10}}

\put(380,24){\makebox(20,12){Meas.}}

\end{picture}
\caption{(a) The procedure for computing one bit of the bit-flip error
syndrome, shown schematically.  The Hadamard gate applied to the ``cat state''
completes the preparation of the Shor state, as discussed in \S3.3.  Both the
XOR gate and the Hadamard gate in the diagram actually represent four gates
performed in parallel. (b) The procedure for computing one bit of the
phase-flip error syndrome, shown schematically. It is the same as (a), but
applied to the data in the Hadamard rotated basis. (c) A circuit equivalent to
(b), simplified by using the identity in Fig.~\ref{identity}.}
\label{fig:synd_circuits}
\end{figure}

Another way to prepare the ancilla was proposed by Andrew Steane.  His 7-qubit
ancilla state is the equally weighted superposition of all Hamming codewords:
\begin{equation}
|{\rm Steane}\rangle_{\rm anc}={1\over 4}\sum_{v\in ~{\rm Hamming}} |v\rangle \
.
\end{equation}
(This state can also be expressed as $\left(|0\rangle_{\rm code} +
|1\rangle_{\rm code}\right)/\sqrt{2}$, and can be obtained by applying the
bitwise Hadamard rotation to
the state $|0\rangle_{\rm code}$.)  To compute the bit-flip syndrome, we
XOR each qubit of the data block into the corresponding qubit of
the ancilla, and measure the ancilla. Applying the Hamming parity-check matrix
$H$ to the {\it classical} measurement outcome, we extract the bit-flip
syndrome. As with Shor's method,  this procedure ``copies'' the
data onto the ancilla, where the state of the ancilla has been carefully chosen
to ensure that only the information about the error can be read by measuring
the ancilla. For example, if there is no error, the particular string that we
find in the measurement is a randomly selected Hamming codeword and tells us
nothing about the state of the data. The same procedure is carried out in the
rotated basis to find
the phase-flip syndrome.  The Steane method has the advantage over the Shor
procedure that only 14 ancilla bits and 14 XOR gates are needed.  But it also
has the disadvantage that the ancilla preparation is more complex, so that the
ancilla is somewhat more prone to error.

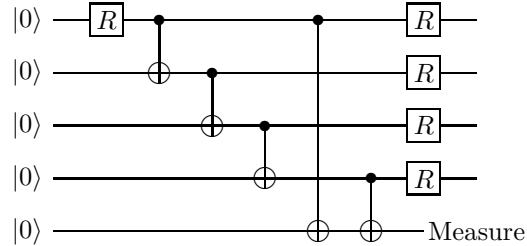
\begin{figure}
\centering
\begin{picture}(210,120)

\put(0,14){\makebox(20,12){$\ket{0}$}}
\put(0,34){\makebox(20,12){$\ket{0}$}}
\put(0,54){\makebox(20,12){$\ket{0}$}}
\put(0,74){\makebox(20,12){$\ket{0}$}}
\put(0,94){\makebox(20,12){$\ket{0}$}}

\put(20,20){\line(1,0){140}}
\put(20,40){\line(1,0){134}}
\put(20,60){\line(1,0){134}}
\put(20,80){\line(1,0){134}}
\put(20,100){\line(1,0){14}}
\put(46,100){\line(1,0){108}}

\put(34,94){\framebox(12,12){$R$}}

\put(60,100){\circle*{4}}
\put(60,100){\line(0,-1){24}}
\put(60,80){\circle{8}}

\put(80,80){\circle*{4}}
\put(80,80){\line(0,-1){24}}
\put(80,60){\circle{8}}

\put(100,60){\circle*{4}}
\put(100,60){\line(0,-1){24}}
\put(100,40){\circle{8}}

\put(120,100){\circle*{4}}
\put(120,100){\line(0,-1){84}}
\put(120,20){\circle{8}}

\put(140,40){\circle*{4}}
\put(140,40){\line(0,-1){24}}
\put(140,20){\circle{8}}

\put(154,34){\framebox(12,12){$R$}}
\put(154,54){\framebox(12,12){$R$}}
\put(154,74){\framebox(12,12){$R$}}
\put(154,94){\framebox(12,12){$R$}}
\put(160,14){\makebox(40,12){Measure}}

\put(166,40){\line(1,0){14}}
\put(166,60){\line(1,0){14}}
\put(166,80){\line(1,0){14}}
\put(166,100){\line(1,0){14}}

\end{picture}
\caption{Construction and verification of the Shor state.  If the measurement
outcome is 1, then the state is discarded and a new Shor state is prepared.}
\label{fig-4qubitcat}
\end{figure}

\subsection{Verifying the Ancilla}

As we continue with our program to sniff our all the ways in which a recovery
failure could result from a single error, we notice another potential problem.
Due to error propagation, a single error that occurs during the preparation of
the Shor state or Steane state could cause two phase errors in this state, and
these can both
propagate to the data if the faulty ancilla is used for syndrome measurement.
Our procedure is not yet fault tolerant.

Therefore the state of the ancilla must be tested for multiple phase errors
before it is
used. If it fails the test, it should be discarded, and a new ancilla state
should be
constructed.

One way to construct and verify the Shor state is shown in
Fig.~\ref{fig-4qubitcat}.  The first Hadamard gate and the first three XOR
gates in this circuit prepare a ``cat state'' $\left( |0000\rangle +
|1111\rangle\right)$, a maximally entangled state of the four ancilla bits; the
final four Hadamard gates rotate the cat state to the Shor state.  But a single
error occuring during the second or third XOR could result in two errors in the
cat state (it might become $\left( |0011\rangle + |1100\rangle\right)$). These
two bit-flip errors in the cat state become two phase errors in the Shor state
which will feed back to cause a block phase error during syndrome measurement.

But we notice that for all the ways that a single bad gate could cause two
bit-flip errors in the cat state, the first and fourth bit of the cat state
will have different values.  Therefore, we add the last two XOR gates to the
circuit (followed by a measurement) to verify whether these two bits of the cat
state agree.  If verification succeeds, we can proceed with syndrome
measurement secure in the knowledge that the probability of two phase errors in
the Shor state is of order $\epsilon^2$.  If verification fails, we can throw
away the cat state and try again.

Of course, a single error in the preparation circuit could also result in two
{\it phase} errors in the cat state and hence two bit-flip errors in the Shor
state; we have made no attempt to check the Shor state for bit-flip errors.
But bit-flip errors in the Shor state are much less troublesome than phase
errors.  Bit-flip errors cause the syndrome measurement to be faulty, but they
do not feed back and damage the data.

If we use Steane's method of syndrome measurement, we first employ the encoding
circuit Fig.~\ref{fig_encode} (with the first two XOR gates
eliminated) to construct $|0\rangle_{\rm code}$, and then apply a Hadamard gate
to each qubit to complete the preparation of the Steane state.  Again, a single
error during encoding can cause two bit flip errors in $|0\rangle_{\rm code}$
which become two phase errors in the Steane state, so that verification is
required. We can verify by performing a nondestructive measurement of the state
to ensure that it is $|0\rangle_{\rm code}$ (up to a single bit flip) rather
than $|1\rangle_{\rm code}$.  Thus we prepare two blocks in the state
$|0\rangle_{\rm code}$, perform a bitwise XOR from the first block to the
second, and then measure the second block destructively.  We can apply
classical Hamming error correction to the measurement outcome, to correct one
possible bit-flip error, and identify the measured block as either
$|0\rangle_{\rm code}$ or $|1\rangle_{\rm code}$.  If the result is
$|0\rangle_{\rm code}$, then the other block has passed inspection.  If the
result is $|1\rangle_{\rm code}$, then we suspect that the other block is
faulty, and we flip that block to fix it.

However, this verification procedure is not yet trustworthy, because it might
have been the block that we measured that was actually faulty, rather than the
block we were trying to check.  Hence we must repeat the verification step.  If
the measured block yields the same result twice in a row, the check may be
deemed reliable.  What if we get a different result the second time?  Then we
don't know whether to flip the block we are checking or to leave it alone.  We
could try one more time, to break the tie, but this is not really necessary; in
fact, if the two verification attempts give conflicting results, it is safe to
do nothing.  Because the results conflict, we know that one of the measured
blocks was faulty.  Therefore, the probability that the block to be checked is
also faulty is order $\epsilon^2$ and can be neglected.  With this verification
procedure, we have managed to construct a Steane state such that the
probability of multiple phase errors (which would feed back to the data during
syndrome measurement) is of order $\epsilon^2$.

\subsection{Verifying the Syndrome}
A single bit-flip error in the ancilla will result in a faulty syndrome.  The
error could arise because the ancilla was prepared incorrectly, or because an
error occured during the syndrome computation.  The latter case is especially
dangerous, because a single error, occuring with a probability of order
$\epsilon$, could produce a fault in {\it both} the data block and the ancilla.
 This might happen because a bad XOR gate causes errors in both its source and
target qubits, or because an error in the data block that occured during
syndrome measurement is later propagated forward to the ancilla by an XOR.

In such cases, were we to accept the faulty syndrome and act to reverse the
error, we would actually introduce a second error into the data block.  So our
procedure is {\it still} not fully fault tolerant; a scenario arising with a
probability of order $\epsilon$ can fatally damage the encoded data.

We must therefore find a way to ensure that the syndrome is more reliable.  The
obvious way to do this is to repeat the syndrome measurement.  It is not
necessary to repeat if the syndrome measurement is trivial (indicates no
error); though there actually might be an error in the data that we failed to
detect, we need not worry that we will make things worse, because if we accept
the syndrome we will take no action.  If on the other hand the syndrome
indicates an error, then we measure the syndrome a second time.  If we obtain
the same result again, it is safe to accept the syndrome and proceed with
recovery, because there is no way occuring with a probability of order
$\epsilon$ to obtain the same (nontrivial) faulty syndrome twice in a row.

If the first two syndrome measurements do not agree, then we could continue to
measure the syndrome until we finally obtain the same result twice in a row, a
result that can be trusted.  Alternatively, we could choose to do nothing,
until an error is reliably detected in a future round of error correction.  At
least in that event we will not make things worse by compounding the error, and
if there is really an error in the data, we will probably detect it next time.

(There are also other ways to increase our confidence in the syndrome.  For
example, instead of repeating the measurement of the entire syndrome, we could
compute some additional redundant syndrome bits, and subject the computed bits
to a parity check.  If there is an error in the syndrome, this method will
usually  detect the error; thus if the parity check passes, the syndrome is
likely to be correct.\cite{evslin,zalka}

Finally, we have assembled all the elements of a fault-tolerant error recovery
procedure. If we take all the precautions described above, then recovery will
fail only if
two independent errors occur, so the probability of an error occurring that
irrevocably damages the encoded block will be of order $\epsilon^2$.

A complete quantum circuit for Steane's error correction is shown in
Fig.~\ref{fig:steane_comp}.  Note that both the bit-flip and phase error
correction are repeated twice.  The verification of the Steane states is also
shown, but the encoding of these states is suppressed in the diagram.
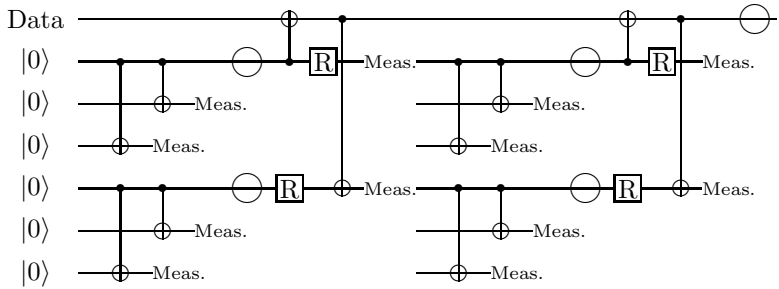
\begin{figure}
\setlength{\unitlength}{.8pt}
\begin{picture}(380,160)

\put(0,14){\makebox(40,12){$|0\rangle$}}
\put(0,34){\makebox(40,12){$|0\rangle$}}
\put(0,54){\makebox(40,12){$|0\rangle$}}
\put(0,74){\makebox(40,12){$|0\rangle$}}
\put(0,94){\makebox(40,12){$|0\rangle$}}
\put(0,114){\makebox(40,12){$|0\rangle$}}
\put(0,134){\makebox(40,12){Data}}

\put(40,20){\line(1,0){35}}
\put(40,40){\line(1,0){55}}
\put(40,60){\line(1,0){94}}
\put(40,80){\line(1,0){35}}
\put(40,100){\line(1,0){55}}
\put(40,120){\line(1,0){110}}

\put(40,140){\line(1,0){330}}

\put(60,60){\circle*{4}}
\put(60,60){\line(0,-1){44}}
\put(60,20){\circle{8}}

\put(60,120){\circle*{4}}
\put(60,120){\line(0,-1){44}}
\put(60,80){\circle{8}}

\put(80,60){\circle*{4}}
\put(80,60){\line(0,-1){24}}
\put(80,40){\circle{8}}

\put(80,120){\circle*{4}}
\put(80,120){\line(0,-1){24}}
\put(80,100){\circle{8}}

\put(75,20){\line(1,1){5}}
\put(75,20){\line(1,-1){5}}
\put(80,14){\makebox(12,12){\scriptsize  ~Meas.}}

\put(75,80){\line(1,1){5}}
\put(75,80){\line(1,-1){5}}
\put(80,74){\makebox(12,12){\scriptsize  ~Meas.}}

\put(95,40){\line(1,1){5}}
\put(95,40){\line(1,-1){5}}
\put(100,34){\makebox(12,12){\scriptsize  ~Meas.}}

\put(95,100){\line(1,1){5}}
\put(95,100){\line(1,-1){5}}
\put(100,94){\makebox(12,12){\scriptsize  ~Meas.}}

\put(120,60){\circle{14}}
\put(120,120){\circle{14}}

\put(140,120){\circle*{4}}
\put(140,120){\line(0,1){24}}
\put(140,140){\circle{8}}

\put(134,54){\framebox(12,12){R}}
\put(150,114){\framebox(12,12){R}}

\put(146,60){\line(1,0){29}}
\put(162,120){\line(1,0){13}}

\put(165,140){\circle*{4}}
\put(165,140){\line(0,-1){84}}
\put(165,60){\circle{8}}

\put(175,60){\line(1,1){5}}
\put(175,60){\line(1,-1){5}}
\put(180,54){\makebox(12,12){\scriptsize  ~Meas.}}

\put(175,120){\line(1,1){5}}
\put(175,120){\line(1,-1){5}}
\put(180,114){\makebox(12,12){\scriptsize  ~Meas.}}

\put(200,20){\line(1,0){35}}
\put(200,40){\line(1,0){55}}
\put(200,60){\line(1,0){94}}
\put(200,80){\line(1,0){35}}
\put(200,100){\line(1,0){55}}
\put(200,120){\line(1,0){110}}

\put(220,60){\circle*{4}}
\put(220,60){\line(0,-1){44}}
\put(220,20){\circle{8}}

\put(220,120){\circle*{4}}
\put(220,120){\line(0,-1){44}}
\put(220,80){\circle{8}}

\put(240,60){\circle*{4}}
\put(240,60){\line(0,-1){24}}
\put(240,40){\circle{8}}

\put(240,120){\circle*{4}}
\put(240,120){\line(0,-1){24}}
\put(240,100){\circle{8}}

\put(235,20){\line(1,1){5}}
\put(235,20){\line(1,-1){5}}
\put(240,14){\makebox(12,12){\scriptsize  ~Meas.}}

\put(235,80){\line(1,1){5}}
\put(235,80){\line(1,-1){5}}
\put(240,74){\makebox(12,12){\scriptsize  ~Meas.}}

\put(255,40){\line(1,1){5}}
\put(255,40){\line(1,-1){5}}
\put(260,34){\makebox(12,12){\scriptsize  ~Meas.}}

\put(255,100){\line(1,1){5}}
\put(255,100){\line(1,-1){5}}
\put(260,94){\makebox(12,12){\scriptsize  ~Meas.}}

\put(280,60){\circle{14}}
\put(280,120){\circle{14}}

\put(300,120){\circle*{4}}
\put(300,120){\line(0,1){24}}
\put(300,140){\circle{8}}

\put(294,54){\framebox(12,12){R}}
\put(310,114){\framebox(12,12){R}}

\put(306,60){\line(1,0){29}}
\put(322,120){\line(1,0){13}}

\put(325,140){\circle*{4}}
\put(325,140){\line(0,-1){84}}
\put(325,60){\circle{8}}

\put(335,60){\line(1,1){5}}
\put(335,60){\line(1,-1){5}}
\put(340,54){\makebox(12,12){\scriptsize  ~Meas.}}

\put(335,120){\line(1,1){5}}
\put(335,120){\line(1,-1){5}}
\put(340,114){\makebox(12,12){\scriptsize  ~Meas.}}

\put(360,140){\circle{14}}

\end{picture}
\caption{The complete circuit for Steane error recovery. Encoded $|0\rangle$'s
are prepared, then verified.  The verified $|0\rangle$'s are used as ancillas
to compute the bit-flip and phase-flip syndromes, which are both measured
twice.  The large circles indicate actions that are taken (conditioned on
measurement outcomes) to repair the ancilla states, or in the final step, to
repair the data block.}
\label{fig:steane_comp}

\end{figure}

\subsection{Measurement and Encoding}
We will of course want to be able to measure our encoded qubits reliably.
But we have already noted in \S2 that {\it destructive}
measurement of the code block is reliable if only one qubit in the block
has a bit-flip error.  If the probability of a flawed measurement is order
$\epsilon$ for a single qubit, then faulty measurements of the code block occur
with probability of order $\epsilon^2$.  Fault-tolerant nondestructive
measurement can also be performed, as we have already noted in our discussion
(\S3.3) of the verification of the Steane state.  An alternative procedure
would be to use the nondestructive measurement depicted in
Fig.~\ref{fig_measure} without any modification.
Though the ancilla is the target of three
successive XOR gates, phase errors feeding back into the block are not so
harmful
because they cannot change $|0\rangle_{\rm code}$ to $|1\rangle_{\rm code}$ (or
vice versa).  However, since a single bit-flip error (in either the data block
or the ancilla qubit) can cause a faulty parity
measurement, the measurement must be repeated ({\it after} bit-flip error
correction) to
ensure accuracy to order $\epsilon^2$.  (We eschewed this procedure in our
description of the verification of the Steane state to avoid the frustration of
needing error correction to prepare the ancilla for error correction!)

We will often want to prepare {\it known} encoded quantum states, such as
$|0\rangle_{\rm code}$.  We already discussed in \S3.3 above (in connection
with preparation of the Steane state), how this encoding can be performed
reliably. In fact, the encoding circuit
is not actually needed.  Whatever the initial state of the block,
(fault-tolerant) error correction will project it onto the space spanned by
$\{|0\rangle_{\rm code},|1\rangle_{\rm code}\}$, and (verified) measurement
will project out either $|0\rangle_{\rm code}$ or $|1\rangle_{\rm code}$.  If
the result $|1\rangle_{\rm code}$ is obtained, then the (bitwise) NOT operator
can be applied to flip the block to the desired state $|0\rangle_{\rm code}$.

If we wish to encode an {\it unknown} quantum state, then we use the encoding
circuit
in Fig.~\ref{fig_encode}.  Again, because of error propagation, a single error
during encoding may cause an encoding failure.  In this case, since no
measurement can verify the encoding, the fidelity of the encoded state will
inevitably be $F=1-O(\epsilon)$.  However, encoding may still be worthwhile,
since it may enable us to preserve the state with a reasonable fidelity for a
longer time than if the state had remained unencoded.

\subsection{Other Codes}

Both Shor's and Steane's scheme for fault-tolerant syndrome measurement have
been described here only for the 7-qubit code, but they can be adapted to more
complex codes that have the capability to recover from many
errors.\cite{divincenzo,steane_c}  Syndrome measurement for more general codes
is best described using the code stabilizer formalism.   In this formalism,
which is discussed in more detail
in the chapter by Andrew Steane in this volume, a quantum error-correcting code
is
characterized as the space of simultaneous eigenstates of a set of commuting
operators (the {\it stabilizer generators}).  Each generator can be expressed
as a product of operators that act on a single qubit, where the single-qubit
operators are chosen from the set $\{I,X,Y,Z\}$ defined in
Eq.~(\ref{xyz_define}).  Each generator squares to the identity and has equal
numbers of eigenvectors with eigenvalue +1 and -1, so that specifying its
eigenvalue reduces the dimension of the space by half.  If there are $n$ qubits
in a block, and there are $n-k$ generators, then the code subspace has
dimension $2^k$ --- there are $k$ encoded qubits.

For example, Steane's 7-qubit code is the space for which the six stabilizer
generators
\begin{eqnarray}
M_1 &=& (IIIZZZZ)\nonumber\\
M_2 &=& (IZZIIZZ)\nonumber\\
M_3 &=& (ZIZIZIZ)\nonumber\\
M_4 &=& (IIIXXXX)\nonumber\\
M_5 &=& (IXXIIXX)\nonumber\\
M_6 &=& (XIXIXIX)\nonumber\\
\end{eqnarray}
all have eigenvalue one.
Comparing to Eq.~(\ref{ham_matrix}), we see that the space with $M_1=M_2=M_3=1$
is spanned by codewords that satisfy the Hamming parity check.  Recalling that
a Hadamard change of basis interchanges $Z$ and $X$, we see that the space with
$M_4=M_5=M_6=1$ is spanned by codewords that satisfy the Hamming parity check
in the Hadamard-rotated basis.  Indeed, the defining property of Steane's code
is that the Hamming parity check is satisfied in both bases.

The stabilizer generators are chosen so that every error operator that is to be
corrected (also expressed as a product of the one-qubit operators
$\{I,X,Y,Z\}$), and the product of any two distinct such error operators,
anticommutes with at least one generator.  Thus, every error changes the
eigenvalues of some of the generators, and two independent errors always change
the eigenvalues in distinct ways.  This means that we obtain a complete error
syndrome by measuring the eigenvalues of all the stabilizer
generators.\footnote{Actually, it is also acceptable if the product of two
independent error operators {\it lies in} the stabilizer.  Then these two
errors will have the same syndrome, but it won't matter, because the two errors
can also be repaired by the same action.  Quantum codes that assign the same
syndrome to more than one error operator are said to be {\it degenerate}.}

Measuring a stabilizer generator $M$ is not difficult.  First we perform an
appropriate unitary change of basis on each qubit so that $M$ in the rotated
basis is a product of $I$'s and $Z$'s acting on the individual qubits. (We
rotate by
\begin{equation}
R={1\over\sqrt{2}}\pmatrix{1&1\cr 1&-1\cr}
\end{equation}
 for each qubit acted on by $X$ in $M$, and by
\begin{equation}
R'={1\over\sqrt{2}}\pmatrix{1&i\cr i& 1\cr}
\end{equation}
 for each qubit acted on by $Y$.) In this basis, the value of $M$ is just the
parity of the bits for which $Z$'s appear. We can measure the parity (much as
we did in our discussion of the 7-qubit code), by performing an XOR to the
ancilla from each qubit in the block for which a $Z$ appears in $M$.  Finally,
we invert the change of basis.  This procedure is repeated for each stabilizer
generator until the entire syndrome is obtained.

We can make this procedure fault-tolerant by preparing the ancilla in a Shor
state for each syndrome bit to be measured, where the number of bits in the
Shor state is the {\it weight} of the corresponding stabilizer generator (the
number of one-qubit operators that are not the identity).  Each ancilla bit is
the target of only a single XOR, so that multiple phase errors do not feed back
into the data.  The procedures discussed above for verifying the Shor state and
the syndrome measurement can also be suitably generalized.

For complex codes that either encode many qubits or can correct many errors,
this generalized Shor method uses many more ancilla qubits and many more
quantum gates than are really necessary to extract an error syndrome.  We can
do considerably better by generalizing the Steane method.  In the case of the
7-qubit code, Steane's idea was that we can use one 7-bit ancilla to measure
all of $M_1$, $M_2$, and $M_3$; we prepare an initial state of the ancilla that
is an equally weighted superposition of all strings that satisfy the Hamming
parity check ({\it i.e}, all words in the classical Hamming code), perform the
appropriate XOR's from the data block to the ancilla, measure all ancilla
qubits, and finally apply the Hamming parity check to the measurement result.
The three parity bits obtained are the measured eigenvalues of $M_1$, $M_2$,
and $M_3$.  The ancilla preparation has been chosen so that no other
information aside from these eigenvalues can be extracted from the measurement
result; hence the coherence of our quantum codewords is not damaged by the
procedure.

This procedure evidently can be adapted to the simultaneous measurement of any
set of operators where each can be expressed as a product of $I$'s and $Z$'s
acting on the individual qubits.  Given a list of $k$ such $n$-qubit operators,
we obtain a matrix $H_Z$ with $k$ rows and $n$ columns by replacing each $I$ in
the list by 0 and each $Z$ by 1.  We prepare the ancilla as the equally
weighted superposition of all length-$n$ strings that obey the $H_Z$ parity
check.  Proceeding with the XOR's and the ancilla measurement (and applying
$H_Z$ to the measurement result), we project a block of $n$ qubits onto a
simultaneous eigenstate of the $n$ operators.  Performing the same procedure in
the Hadamard-rotated basis, we can simultaneously measure any set of operators
where each is a product of $I$'s and $X$'s.

Among the stabilizer generators there also might be operators that have the
form $M=\bar Z\bar X$, where $\bar Z$ is a product of $Z$'s acting on one set
of qubits, and $\bar X$ is a product of $X$'s acting on another set of qubits.
Since the generator $M$ must square to the identity, the number of qubits acted
on by the product $Y$ of $Z$ and $X$ must be even. Hence $\bar Z$ and $\bar X$
commute, and so can be simultaneously measured by the method described above.
However, this measurement would give too much information; we want to measure
the product of $\bar Z$ and $\bar X$ rather than measure each separately.  To
make the measurement we want, we must further modify the ancilla.  The ancilla
should not be chosen to satisfy both the $H_Z$ parity check and the
corresponding $H_X$ parity check. Rather it is prepared so that the $H_Z$ and
$H_X$ parity bits are correlated --- the ancilla is a sum over strings such
that either both parity bits are trivial or both bits are nontrivial.  After
the ancilla measurement, we sum the parity of the ``$\bar Z$ measurement'' and
the ``$\bar X$ measurement'' to obtain the eigenvalue of $M$.  But the separate
parities of the $\bar Z$ and $\bar X$ ``measurements'' are entirely random and
actually reveal nothing about the values of $\bar Z$ or $\bar X$.

Now we can describe Steane's method in its general form that can be applied to
any stabilizer code.  If $k$ logical qubits are encoded in a block of $n$
qubits, then there are $n-k$ independent stabilizer generators.  With a list of
these generators we associate a matrix
\begin{equation}
\bar H=\pmatrix{H_Z &|& H_X\cr}
\end{equation}
that has $n-k$ rows and $2n$ columns.  The positions of the 1's in $H_Z$
indicate the qubits that are acted on by $Z$ in the listed generators, and the
1's in $H_X$ indicate the qubits acted on by $X$; if a 1 appears in the same
position in both $H_Z$ and $H_X$, then the product $Y=ZX$ acts on that qubit.
A $2n$-qubit ancilla is prepared in the {\it generalized Steane state} --- the
equally weighted superposition of all of the strings that satisfy the $\bar H$
parity check.  Then the quantum circuit shown in Fig.~\ref{fig:steane_circ} is
executed, the ancilla qubits are measured, and $\bar H$ is applied to the
measurement result.  The parity bits found are the eigenvalues of the
stabilizer generators, which provide the complete error syndrome.

\begin{figure}
\centering
\begin{picture}(180,85)

\put(0,14){\makebox(20,12){Steane}}
\put(0,74){\makebox(20,12){Data}}

\put(30,10){\line(1,0){10}}
\put(40,4){\framebox(12,12){$R$}}
\put(52,10){\line(1,0){56}}
\put(108,4){\framebox(12,12){$R$}}
\put(120,10){\line(1,0){10}}

\put(30,30){\line(1,0){100}}

\put(30,80){\line(1,0){100}}

\put(60,80){\circle*{6}}
\put(60,80){\line(0,-1){55}}
\put(60,30){\circle{10}}

\put(80,10){\circle*{6}}
\put(80,10){\line(0,1){75}}
\put(80,80){\circle{10}}

\put(140,14){\makebox(20,12){Measure}}

\end{picture}
\caption{Circuit for Steane syndrome measurement, shown schematically. A
$2n$-qubit Steane state is used to find the syndrome for an $n$-qubit data
block. Each XOR gate in the diagram represents $n$ XOR gates performed in
parallel.}
\label{fig:steane_circ}
\end{figure}
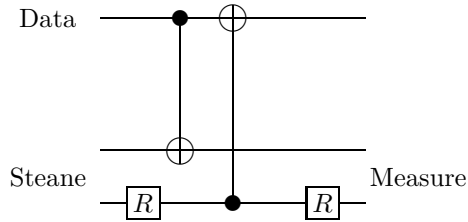

The ancilla preparation has been designed so that no other information other
than the syndrome can be extracted from the measurement result, and therefore
the coherence of the quantum codewords is not damaged by the procedure.  Each
qubit in the code block is acted on by only two quantum gates in this
procedure, the minimum necessary to detect both bit-flip and phase errors
afflicting any qubit.

Finally, we note that a different strategy for performing fault-tolerant error
correction was described by Kitaev.\cite{kitaev_a}  He invented a family of
quantum error-correcting codes such that many errors within the code block can
be corrected, but only four XOR gates are needed to compute each bit of the
syndrome.  In this case, even if we use just a single ancilla qubit for the
computation of each syndrome bit (rather than an expanded ancilla state like a
Shor or Steane state), only a limited number of errors can feed back from the
ancilla into the data.  The code can then be chosen such that the typical
number of errors fed back into the data during the syndrome computation is
comfortably less than the maximum number of errors that the code can tolerate.

\section{Fault-tolerant quantum gates}
\label{sec:gates}

We have seen that coding can protect quantum information. But we want to do
more than {\it store} quantum information with high fidelity; we want to
operate a quantum computer that {\it processes} the information.   Of course,
we could decode, perform a gate, and then re-encode, but that procedure would
temporarily expose the quantum information to harm.  Instead, if we want our
quantum computer to operate reliably, we must be able to apply quantum gates
directly to the encoded data, and these gates must respect the principles of
fault tolerance if catastrophic propagation of error is to be avoided.

\subsection{The 7-qubit code}
In fact, with Steane's 7-qubit code, there are a number of gates that can be
easily implemented.  Three single-qubit gates can all be applied {\it bitwise};
that is applying these gates to each of the 7 qubits in the block implements
the same gate acting on the encoded qubit.  We have already seen in
Eq.~(\ref{tildezero}) that the Hadamard rotation $R$ acts this way.  The same
is
true for the NOT gate (since each odd parity Hamming codeword is the complement
of an even parity Hamming codeword)\footnote{Actually, we can implement the NOT
acting on the encoded qubit with just 3 NOT's applied to selected qubits in the
block.}, and the phase gate
\begin{equation}
P=\pmatrix{1&0\cr 0& i\cr} \ ;
\end{equation}
(the odd Hamming codewords have weight $\equiv$ 3 (mod 4) and the even
codewords have weight $\equiv$ 0 (mod 4), so we actually apply $P^{-1}$ bitwise
to implement $P$).  The XOR gate can also be implemented bitwise; that is, by
XOR'ing each bit of the source block into the corresponding bit of the target
block, as in Fig.~\ref{fig:transversal}.  This works because the even codewords
form a subcode, while the odd
codewords are its nontrivial coset.

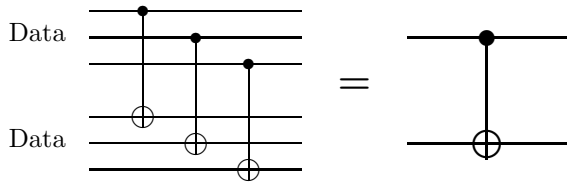
\begin{figure}
\centering
\begin{picture}(210,85)

\put(0,26){\makebox(20,12){Data}}
\put(0,66){\makebox(20,12){Data}}

\put(30,20){\line(1,0){80}}
\put(30,30){\line(1,0){80}}
\put(30,40){\line(1,0){80}}
\put(30,60){\line(1,0){80}}
\put(30,70){\line(1,0){80}}
\put(30,80){\line(1,0){80}}

\put(50,80){\circle*{4}}
\put(50,80){\line(0,-1){44}}
\put(50,40){\circle{8}}

\put(70,70){\circle*{4}}
\put(70,70){\line(0,-1){44}}
\put(70,30){\circle{8}}

\put(90,60){\circle*{4}}
\put(90,60){\line(0,-1){44}}
\put(90,20){\circle{8}}

\put(120,46){\makebox(20,12){\Large\bf =}}

{\thicklines
\put(150,30){\line(1,0){60}}
\put(150,70){\line(1,0){60}}

\put(180,70){\circle*{6}}
\put(180,70){\line(0,-1){46}}
\put(180,30){\circle{10}}
}

\end{picture}
\caption{The transversal XOR gate, shown schematically. By XOR'ing each bit of
the source block into the corresponding bit of the target block, we implement
an XOR acting on the encoded qubits. The gate implementation is fault tolerant
because each qubit in both code blocks is acted on by a single gate.}
\label{fig:transversal}
\end{figure}

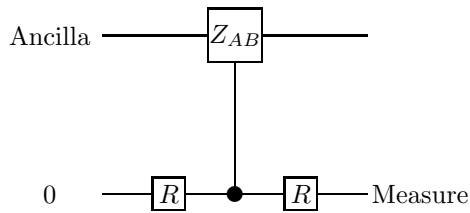
\begin{figure}
\centering
\begin{picture}(180,85)

\put(0,14){\makebox(20,12){0}}
\put(0,74){\makebox(20,12){Ancilla}}

\put(30,20){\line(1,0){19}}
\put(49,14){\framebox(12,12){$R$}}
\put(61,20){\line(1,0){38}}
\put(99,14){\framebox(12,12){$R$}}
\put(111,20){\line(1,0){19}}

\put(30,80){\line(1,0){40}}
\put(70,70){\framebox(20,20){$Z_{AB}$}}
\put(90,80){\line(1,0){40}}

\put(80,20){\circle*{6}}
\put(80,20){\line(0,1){50}}

\put(140,14){\makebox(20,12){Measure}}

\end{picture}
\caption{The measurement circuit used in the ancilla preparation step of Shor's
implementation of the Toffoli gate.}
\label{fig:shor_meas}
\end{figure}

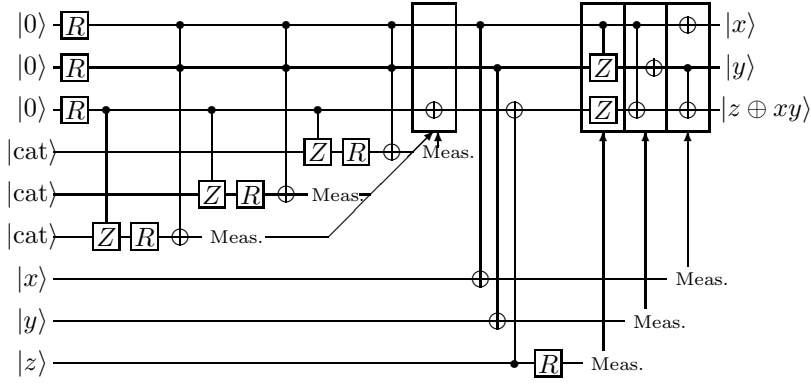
\begin{figure}
\centering
\setlength{\unitlength}{.8pt}
\begin{picture}(335,200)

\put(0,174){\makebox(20,12){$\ket{0}$}}
\put(0,154){\makebox(20,12){$\ket{0}$}}
\put(0,134){\makebox(20,12){$\ket{0}$}}
\put(0,114){\makebox(20,12){$\ket{\rm cat}$}}
\put(0,94){\makebox(20,12){$\ket{\rm cat}$}}
\put(0,74){\makebox(20,12){$\ket{\rm cat}$}}
\put(0,54){\makebox(20,12){$\ket{x}$}}
\put(0,34){\makebox(20,12){$\ket{y}$}}
\put(0,14){\makebox(20,12){$\ket{z}$}}

\put(20,180){\line(1,0){4}}
\put(24,174){\framebox(12,12){$R$}}
\put(20,160){\line(1,0){4}}
\put(24,154){\framebox(12,12){$R$}}
\put(20,140){\line(1,0){4}}
\put(24,134){\framebox(12,12){$R$}}

\put(36,180){\line(1,0){299}}
\put(36,160){\line(1,0){238}}
\put(286,160){\line(1,0){49}}
\put(36,140){\line(1,0){238}}
\put(286,140){\line(1,0){49}}

\put(20,80){\line(1,0){19}}
\put(51,80){\line(1,0){6}}
\put(69,80){\line(1,0){21}}
\put(57,74){\framebox(12,12){$R$}}

\put(39,74){\framebox(12,12){$Z$}}
\put(45,140){\circle*{4}}
\put(45,140){\line(0,-1){54}}

\put(80,180){\circle*{4}}
\put(80,180){\line(0,-1){104}}
\put(80,160){\circle*{4}}
\put(80,80){\circle{8}}

\put(90,74){\makebox(30,12){\scriptsize  ~Meas.}}

\put(20,100){\line(1,0){69}}
\put(101,100){\line(1,0){6}}
\put(119,100){\line(1,0){21}}
\put(107,94){\framebox(12,12){$R$}}

\put(89,94){\framebox(12,12){$Z$}}
\put(95,140){\circle*{4}}
\put(95,140){\line(0,-1){34}}

\put(130,180){\circle*{4}}
\put(130,180){\line(0,-1){84}}
\put(130,160){\circle*{4}}
\put(130,100){\circle{8}}

\put(138,94){\makebox(30,12){\scriptsize  ~Meas.}}

\put(20,120){\line(1,0){119}}
\put(151,120){\line(1,0){6}}
\put(169,120){\line(1,0){21}}
\put(157,114){\framebox(12,12){$R$}}

\put(139,114){\framebox(12,12){$Z$}}
\put(145,140){\circle*{4}}
\put(145,140){\line(0,-1){14}}

\put(180,180){\circle*{4}}
\put(180,180){\line(0,-1){64}}
\put(180,160){\circle*{4}}
\put(180,120){\circle{8}}

\put(190,114){\makebox(30,12){\scriptsize  ~Meas.}}

\put(200,140){\circle{8}}
\put(200,136){\line(0,1){8}}
\put(190,130){\framebox(20,60){}}
\put(150,80){\vector(1,1){50}}
\put(120,80){\line(1,0){30}}
\put(165,100){\line(1,0){5}}
\put(202,122){\vector(0,1){8}}

\put(20,60){\line(1,0){290}}
\put(20,40){\line(1,0){270}}
\put(20,20){\line(1,0){228}}

\put(222,180){\circle*{4}}
\put(222,180){\line(0,-1){124}}
\put(222,60){\circle{8}}

\put(230,160){\circle*{4}}
\put(230,160){\line(0,-1){124}}
\put(230,40){\circle{8}}

\put(238,20){\circle*{4}}
\put(238,20){\line(0,1){124}}
\put(238,140){\circle{8}}

\put(248,14){\framebox(12,12){$R$}}
\put(260,20){\line(1,0){10}}

\put(270,14){\makebox(30,12){\scriptsize  ~Meas.}}
\put(290,34){\makebox(30,12){\scriptsize  ~Meas.}}
\put(310,54){\makebox(30,12){\scriptsize  ~Meas.}}

\put(280,180){\circle*{4}}
\put(280,180){\line(0,-1){14}}
\put(274,154){\framebox(12,12){$Z$}}
\put(274,134){\framebox(12,12){$Z$}}
\put(270,130){\framebox(20,60){}}
\put(280,26){\vector(0,1){104}}

\put(296,180){\circle*{4}}
\put(296,180){\line(0,-1){44}}
\put(296,140){\circle{8}}
\put(304,160){\circle{8}}
\put(304,156){\line(0,1){8}}
\put(290,130){\framebox(20,60){}}
\put(300,46){\vector(0,1){84}}

\put(320,180){\circle{8}}
\put(320,176){\line(0,1){8}}
\put(320,160){\circle*{4}}
\put(320,160){\line(0,-1){24}}
\put(320,140){\circle{8}}
\put(310,130){\framebox(20,60){}}
\put(320,66){\vector(0,1){64}}

\put(335,174){\makebox(20,12){$\ket{x}$}}
\put(335,154){\makebox(20,12){$\ket{y}$}}
\put(335,134){\makebox(40,12){$~\ket{z \oplus xy}$}}

\end{picture}
\caption[The fault-tolerant Toffoli gate.]{The fault-tolerant Toffoli gate.
Each 
line represents a block of 7 qubits, and the gates are implemented
transversally. For each measurement, the arrow points to the set of gates that
is to be applied if the measurement outcome is 1; no action is taken if the
outcome is 0.}
\label{fig-toffoli}
\end{figure}

Thus there are simple fault-tolerant procedures for implementing the gates NOT,
$R$, $P$, and XOR.  But unfortunately, these gates do not by themselves form a
universal set.  To be able to perform any desired unitary transformation acting
on
encoded data (to arbitrary precision), we will need to make a suitable addition
to this set.  Following
Shor,\cite{shor_c} we will add the 3-qubit Toffoli gate, which is implemented
by the
procedure shown in Fig.~\ref{fig-toffoli}.\footnote{Knill {\it et
al.}\cite{knill_c,knill_d} describe an alternative way of completing the
universal set of gates.}  

Shor's construction of the fault-tolerant Toffoli gate has two stages.  In the
first stage, three encoded ancilla blocks
are prepared in a state of the form
\begin{equation}
\label{A_define}
|A\rangle_{\rm anc}\equiv {1\over {2}}\sum_{a=0,1}\sum_{b=0,1}
|a,b,ab\rangle_{\rm anc} \
.
\end{equation}
In the second stage, the ancilla interacts with three data blocks to complete
the execution of the gate.  First, we will describe how the ancilla is
prepared.  To begin with, each of three ancilla blocks are encoded in the state
$|0\rangle_{\rm code}$.  Bitwise Hadamard gates are applied to all three blocks
to prepare the encoded state
\begin{equation}
{1\over\sqrt{8}}\sum_{a=0,1}\sum_{b=0,1} \sum_{c=0,1}|a,b,c\rangle_{\rm anc} \
.
\end{equation}
We note that this state can be expressed as
\begin{equation}
\label{not3}
{1\over \sqrt{2}}\left(|A\rangle_{\rm anc}+ |B\rangle_{\rm anc}\right) \ ,
\quad |B\rangle_{\rm anc}\equiv
{\rm NOT}_3|A\rangle_{\rm anc} \ ,
\end{equation}
where ${\rm NOT}_3$ denotes a NOT gate acting on the third encoded qubit.  In
the remainder of the ancilla preparation, the three blocks are measured in the
$\{|A\rangle_{\rm anc},|B\rangle_{\rm anc}\}$ basis; if the outcome
$|A\rangle_{\rm anc}$  is obtained, the preparation is complete; if
$|B\rangle_{\rm anc}$ is obtained, ${\rm NOT}_3$ is applied to complete the
procedure.

Now we must explain how the $\{|A\rangle_{\rm anc},|B\rangle_{\rm anc}\}$
measurement is carried out.  Schematically, the measurement is done with the
circuit shown in Fig.~\ref{fig:shor_meas}, where the $Z_{AB}$ gate (conditioned
on a control bit) flips the relative phase of $|A\rangle_{\rm anc}$ and
$|B\rangle_{\rm anc}$.  We can see from Eqs.~(\ref{A_define}) and (\ref{not3})
that, in terms of the values $a$, $b$, and $c$ of the three ancilla blocks,
$Z_{AB}$ applies the phase $(-1)^{ab+c}$.  If the control bit is denoted $x$,
then the gates we need to apply are $(-1)^{xab}$ and $(-1)^{xc}$, the product
of a three-bit phase gate and a two-bit phase gate.

But a three-bit phase gate is as hard as a Toffoli gate, so we seem to be
stuck.  However, we can get around this impasse if the control block is chosen
to be not an encoded qubit, but instead a (verified) 7-bit ``cat
state'' 
\begin{equation}
|{\rm cat}\rangle={1\over\sqrt{2}}\bigl(|0000000\rangle + |1111111\rangle\bigr)
\ .
\end{equation}
We {\it do} already know how to construct fault-tolerant {\it two-bit} and {\it
one-bit} phase gates transversally.  These can be promoted to the three-bit and
two-bit gates that we need by simply conditioning all of the bitwise gates in
the construction on the corresponding bits of the cat state.  Finally, we apply
the bitwise Hadamard rotation to the cat state and measure its parity to
complete the execution of the measurement circuit Fig.~\ref{fig:shor_meas}.
(We obtain the circuit in Fig.~\ref{fig-toffoli}, by noting that, if the cat
state is in the Hadamard rotated basis, the three-bit phase gate can be
expressed as a Toffoli gate with the cat state as target; therefore one bitwise
Toffoli gate is executed in our implementation of the measurement circuit.)  Of
course, the measurement is repeated to ensure accuracy.

Meanwhile, three data blocks have been waiting patiently for the ancilla to be
ready.  By applying three XOR gates and a Hadamard rotation, the state of the
data and ancilla is transformed as 
\begin{eqnarray}
&&\sum_{a=0,1}\sum_{b=0,1} |a,b,ab\rangle_{\rm anc}|x,y,z\rangle_{\rm
data}\nonumber\\
&&\longrightarrow
\sum_{a=0,1}\sum_{b=0,1}\sum_{w=0,1} (-1)^{wz} |a,b,ab\oplus z\rangle_{\rm
anc}|x\oplus a,y\oplus b,w\rangle_{\rm data} \ . \\
\end{eqnarray}
Now each {\it data} block is measured.  If the measurement outcome is 0, no
action is taken, but if the measurement outcome is 1, then a particular set of
gates is applied to the {\it ancilla}, as shown in Fig.~\ref{fig-toffoli}, to
complete the implementation of the Toffoli gate.  Note that the original data
blocks are destroyed by the procedure, and that what were initially the ancilla
blocks become the new data blocks. The important thing about this construction
is that all of the steps have been carefully designed to adhere to the
principles of
fault tolerance and minimize the propagation of error.  Thus, two independent
errors must occur during the procedure in order for two errors to arise in any
one of the data blocks. 

That the fault-tolerant gates form a discrete set is a bit of a nuisance, but
it is also an unavoidable feature of any fault-tolerant scheme.  It would not
make sense for the fault-tolerant gates to form a continuum, for then how could
we possibly avoid making an error by applying the {\it wrong} gate, a gate that
differs from the intended one by a small amount?  Anyway, since our
fault-tolerant gates form a universal set,  they suffice for approximating any
desired unitary transformation to any desired accuracy.

\subsection{Other codes}
Shor\cite{shor_c} described how to generalize this fault tolerant set of gates
to more
complex codes that can correct more errors, and
Gottesman\cite{gottesman_b,gottesman_c} has
described an even more general procedure that can be applied to any of the
quantum stabilizer codes.

Gottesman's construction begins with the observation that for any stabilizer
code, there are fault-tolerant implementations of the single qubit gates $X$
and $Z$ acting on each encoded qubit.  For a stabilizer code with block size
$n$, recall that we have already seen in \S3.6 that any ``error operator'' $M$
(expressed as a tensor product of $n$ matrices chosen from $\{I,X,Y,Z\}$) can
be written in the form $\bar Z\bar X$, and so can be uniquely represented as a
binary string of length $2n$. If there are $k$ logical qubits encoded in the
block, then the stabilizer of the code is generated by $n-k$ such operators.
The error operators that commute with all elements of the stabilizer themselves
form a group. The generators of this group are represented by binary strings of
length $2n$ that are required to satisfy $n-k$ independent binary conditions;
therefore, there are $n+k$ independent generators.  Of these, $n-k$ are the
generators of the stabilizer, but there are in addition $2k$ independent error
operators that do not lie in the stabilizer, but do commute with the
stabilizer.  These $2k$ operators preserve the code subspace but act
nontrivially on the codewords, and therefore they can be interpreted as
operations that act on the encoded logical qubits.

In fact, these $2k$ operators can be chosen as the single qubit operations
$\hat Z_i$ and $\hat X_i$, where $i=1,2,3,\dots,k$ labels the encoded qubits.
We first note that the $n-k$ stabilizer generators can be extended to a maximal
commuting set of $n$ operators; the $k$ additional operators may be identified
as the $\hat Z_i$'s.  We can choose the computational basis states in the code
subspace to be the simultaneous eigenstates of all the $\hat Z_i$'s, with the
$+1$ eigenvalue corresponding to the value $0$, and the $-1$ eigenvalue to the
value $1$.  Then $\hat Z_i$ flips the phase of qubit $i$.  We may choose the
remaining $k$ generators, denoted $\hat X_i$, which commute with the stabilizer
but not with all of the $\hat Z_i$'s, to obey the relations
\begin{equation}
\label{zx_form}
[\hat Z_i,\hat Z_j]=0=[\hat X_i,\hat X_j] \ ,\quad [\hat Z_i, \hat X_j]=0 \ ,
(i\ne j) \ , \quad \hat Z_i \hat X_i + \hat X_i \hat Z_i = 0 \ .
\end{equation}
Since $\hat X_i$ anticommutes with $\hat Z_i$, it flips the eigenvalue of $\hat
Z_i$, and hence the value of qubit $i$.  All of these operations are performed
by applying at most one single-qubit gate to each qubit in the block;
therefore, these operations are surely fault tolerant.

We have also seen in \S3.6 that it is possible to perform a fault-tolerant {\it
measurement} of {\it any} error operator $\bar Z\bar X$, and so in particular
to measure each $\hat X_i$, $\hat Y_i$, and $\hat Z_i$ fault tolerantly.
Gottesman\cite{gottesman_b} has shown that, if it possible to perform a Toffoli
gate (which is universal for the {\it classical} computations that preserve the
set of computational basis states), then the single qubit gates $X$ and $Z$,
together with the ability to measure $X$, $Y$, and $Z$ for any qubit, suffice
for universal quantum computation.  Hence, if we can show that a fault-tolerant
Toffoli gate can be constructed acting on any three qubits, we will have
completed the demonstration that universal fault-tolerant quantum computation
is possible with any stabilizer code.

The construction of a fault-tolerant Toffoli gate is rather complicated, so it
is best to organize the demonstration this way:  Gottesman showed that in any
stabilizer code, it is possible to apply a fault-tolerant XOR gate to any pair
of qubits (whether or not the two qubits reside in the same code block).  Using
the XOR gate, plus the single qubit gates and measurements that we have already
seen can be implemented fault-tolerantly, one can show that all of the gates
needed in Shor's construction of the Toffoli gate can be constructed.  Thus,
the fault-tolerant construction of the Toffoli gate can be carried out using
any stabilizer code, and universal fault-tolerant quantum computation can be
achieved.

While in principle any stabilizer code can be used for fault-tolerant quantum
computing, some codes are better than
others.  For example, there is a 5-qubit code that can recover from one
error\cite{bennett,laflamme}
 and Gottesman\cite{gottesman_b} has
exhibited a universal set of fault-tolerant gates for this code.  But the gate
implementation is quite complex.  The 7-qubit Steane code requires a larger
block, but it is much more convenient for computation.

\section{The accuracy threshold for quantum computation}

Quantum error-correcting codes exist that can correct $t$ errors, where $t$ can
be arbitrarily large.  If we use such a code and we follow the principles of
fault-tolerance, then an uncorrectable error will occur only if at least $t+1$
{\it independent} errors occur in a single block before recovery is completed.
So if the probability of an error occurring per quantum gate, or the
probability of a storage error occurring per unit of time, is of order
$\epsilon$, then the probability of an error per gate acting on encoded data
will be of order $\epsilon^{t+1}$, which is much smaller if $\epsilon$ is small
enough.  Indeed, it may seem that by choosing a code with $t$ as large as we
please we can make the probability of error per gate as small as we please, but
this turns out not to be the case, at least not for most codes.  The trouble is
that as we increase $t$, the complexity of the code increases sharply, and the
complexity of the recovery procedure correspondingly increases.  Eventually we
reach the point where it takes so long to perform recovery that it is likely
that $t+1$ errors will accumulate in a block before we can complete the
recovery step, and the ability of the code to correct errors is thus
compromised.

Suppose that the number of computational steps needed to perform the syndrome
measurement increases with $t$ like a power $t^b$.  Then the probability that
$t+1$ errors accumulate before the measurement is complete will behave like
\begin{equation}
\label{t_prob}
{\rm Block ~Error ~Probability}\sim \left( t^b \epsilon\right)^{t+1}\  ,
\end{equation}
where $\epsilon$ is the probability of error per step.
We may then choose $t$ to minimize the error probability ($t\sim
e^{-1}\epsilon^{-1/b}$, assuming $t$ is large), obtaining
\begin{equation}
{\rm Minimum ~Block ~Error ~Probability}\sim
\exp\left(-e^{-1}b\epsilon^{-1/b}\right) \ .
\end{equation}
Thus if we hope to carry out altogether $T$ cycles of error correction without
any error occurring, 
then our gates must have an accuracy
\begin{equation}
\label{acc_need}
\epsilon\sim \left(\log T\right)^{-b} \ .
\end{equation}
Similarly, if we hope to perform a quantum computation with altogether $T$
quantum gates, elementary gates of this prescribed accuracy are needed.

In the procedure originally described by Shor,\cite{shor_c} the power
characterizing
the complexity of the syndrome measurement is $b=4$, and somewhat smaller
values of
$b$ can be achieved with a  better optimized procedure.   The block
size of the code used grows with $t$ like $t^2$ (for the codes that Shor
considered), so when the code is chosen to
optimize the error probability, the block size is of order $(\log T)^2$.
Certainly, the scaling described by  Eq.~(\ref{acc_need}) is much more
favorable
than the accuracy $\epsilon\sim T^{-1}$ that would be required were coding not
used at all.  But for any given accuracy, there is a limit to how long a
computation can proceed until errors become likely.

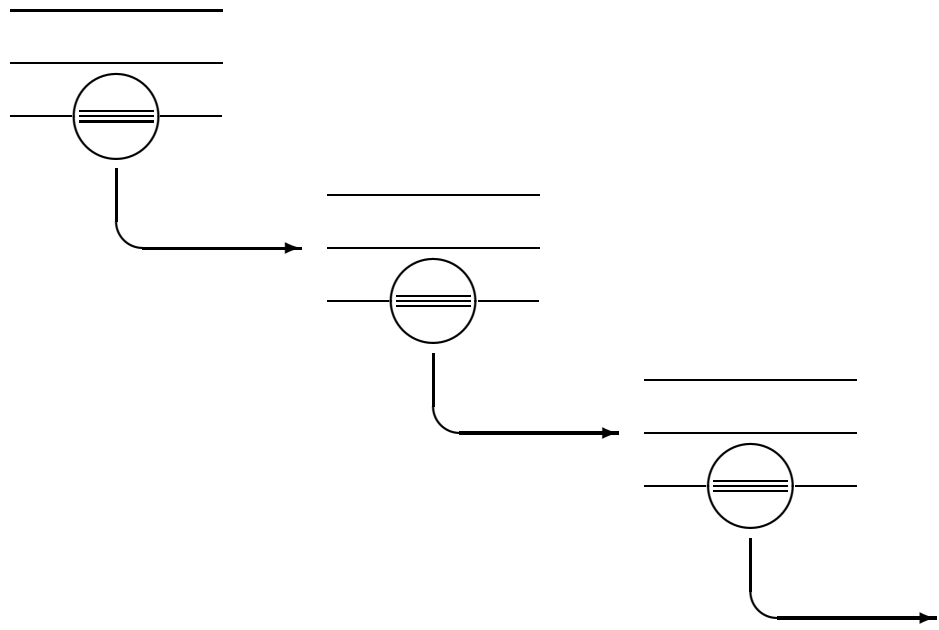
\begin{figure}
\centering
\begin{picture}(360,230)

\put(0,230){\line(1,0){80}}
\put(0,210){\line(1,0){80}}
\put(0,190){\line(1,0){23}}
\put(57,190){\line(1,0){23}}

{\thicklines
\put(40,190){\circle{30}}
}
\put(26,192){\line(1,0){28}}
\put(26,190){\line(1,0){28}}
\put(26,188){\line(1,0){28}}

{\thicklines
\put(40,170){\line(0,-1){20}}
\put(50,150){\oval(20,20)[bl]}
\put(50,140){\vector(1,0){60}}
}

\put(120,160){\line(1,0){80}}
\put(120,140){\line(1,0){80}}
\put(120,120){\line(1,0){23}}
\put(177,120){\line(1,0){23}}

{\thicklines
\put(160,120){\circle{30}}
}

\put(146,122){\line(1,0){28}}
\put(146,120){\line(1,0){28}}
\put(146,118){\line(1,0){28}}

{\thicklines
\put(160,100){\line(0,-1){20}}
\put(170,80){\oval(20,20)[bl]}
\put(170,70){\vector(1,0){60}}
}

\put(240,90){\line(1,0){80}}
\put(240,70){\line(1,0){80}}
\put(240,50){\line(1,0){23}}
\put(297,50){\line(1,0){23}}

{\thicklines
\put(280,50){\circle{30}}
}

\put(266,48){\line(1,0){28}}
\put(266,50){\line(1,0){28}}
\put(266,52){\line(1,0){28}}

{\thicklines
\put(280,30){\line(0,-1){20}}
\put(290,10){\oval(20,20)[bl]}
\put(290,0){\vector(1,0){60}}
}

\end{picture}
\caption{Concatenated coding.  Each qubit in the block, when inspected at
higher resolution, is itself an encoded subblock.}
\label{fig_concatenate}
\end{figure}

This limitation can be overcome by using a special kind of code, a {\it
concatenated} code.\cite{knill_a}$^{\!-\,}$\cite{zalka} To understand the
concept of a
concatenated code, imagine that we are using Steane's quantum error-correcting
code that encodes a single qubit as a block of $7$ qubits.  But if we look more
closely at one of the $7$ qubits in the block with better resolution, we
discover that it is not really a single qubit, but another block of $7$ encoded
using the same Steane code as before.  And when we examine one of the $7$
qubits in {\it this} block with higher resolution, we discover that it too is
really a block of $7$ qubits.  And so on. (See Fig.~\ref{fig_concatenate}.) If
there are all together $L$ levels to this hierarchy of concatenation, then a
single qubit is actually encoded in a block of size $7^L$.  The reason that
concatenation is useful is that it enables us to recover from errors more
efficiently, by dividing and conquering. With this method, the
complexity of error correction does not grow so sharply as we increase the
error-correcting capacity of our quantum code.

We have seen that Steane's 7-qubit code can recover from one error.  If the
probability of error per qubit is $\epsilon$, the errors are uncorrelated, and
recovery is fault-tolerant, then the probability of a recovery failure is of
order $\epsilon^2$.  If we concatenate the code to construct a block of size
$7^2$, then an error occurs in the block only if two of the subblocks of size 7
fail, which occurs with a probability of order $(\epsilon^2)^2$.  And if we
concatenate again, then an error occurs only if two subblocks of size $7^2$
fail, which occurs with a probability of order $((\epsilon^2)^2)^2$.  If there
are all together $L$ levels of concatenation, then the probability of an error
is or order $(\epsilon)^{2^L}$, while the block size is $7^L$.  Now, if the
error rate for our fundamental gates is small enough, then we can improve the
probability of an error per gate by concatenating the code.  If so, we can
improve the performance even more by adding another level of concatenation.
And so on.    This is the origin of the accuracy threshold for  quantum
computation:  if coding reduces the probability of error significantly, then we
can make the error rate arbitrarily small by adding enough levels of
concatenation.  But if the error rates are too high to begin with, then coding
will make things worse instead of better.

To analyze this situation, we must first adopt a particular model of the
errors,  and I will choose the simplest possible quasi-realistic model:
uncorrelated stochastic errors.\footnote{I will characterize the error model in
more detail in \S\ref{dream}.}  In  each
computational time step, each qubit in the device becomes entangled with the
environment as in Eq.~(\ref{superop}), but where we assume that the four states
of the environment are mutually orthogonal, and that the three ``error states''
have equal norms.  Thus the three types of errors (bit flip, phase flip, both)
are assumed to be equally likely.  The total probability of error in each time
step is denoted $\epsilon_{\rm store}$.  In addition to these storage errors
that afflict the ``resting'' qubits, there will also be errors that are
introduced  by the quantum gates themselves.
For each type of gate, the probability of error each time the gate is
implemented is denoted $\epsilon_{\rm gate}$ (with independent values assigned
to gates of each type).  If the gate acts on more than one qubit (XOR or
Toffoli), correlated errors may arise. In our analysis, we make the pessimistic
assumption
that an error in a multi-qubit gate always damages all of the qubits on which
the gate acts;
{\it e.g.}, a faulty XOR gate introduces errors in both the source qubit and
the target qubit. This assumption (among others) is made only to keep the
analysis tractable.  Under more realistic assumptions, we would find that
somewhat higher error rates could be tolerated.

We can analyze the efficacy of concatenated coding by constructing a set of
{\it concatenation flow equations}, that describe how our error model evolves
as we proceed from one level of concatenation to the next.  For example,
suppose we want to perform an XOR gate followed by an error recovery step on
qubits encoded using the concatenated Steane code with $L$ levels of
concatenation (block size $7^L$).   The gate implementation and recovery can be
described in terms of operations that act on subblocks of size $7^{L-1}$.
Thus, we can derive an expression for the probability of error $\epsilon^{(L)}$
for a gate acting on the full block in terms of the probability of error for
gates acting on the subblocks.  This expression is one of the flow equations.
In principle, we can solve the flow equations to find the error probabilities
``at level $L$'' in terms of the parameters of the error model for the
elementary qubits. We then study how the error probabilities behave as $L$
becomes large.  If all block error probabilities approach zero for $L$ large,
then the elementary error probabilities are ``below the threshold.''  Since our
elementary error model may have many parameters, the threshold is really a
codimension one surface in a high-dimension space.

Steane's method of syndrome measurement is particularly well suited for
concatenated coding.  All of the gates in the recovery circuit
Fig.~\ref{fig:steane_comp} can be executed {\it transversally}; if we perform
the gates on the elementary qubits in the code block, then we are executing the
very same gates on the encoded information in each block of size 7, each
superblock of size $7\cdot 7$ and so on. Similarly, when we measure the
elementary qubits in the ancilla at the conclusion of the syndrome computation,
then (after applying classical Hamming error correction to the qubits), we have
also measured the encoded qubit in each block of 7, and (after applying Hamming
error correction to the blocks) each superblock of $7\cdot 7$, {\it etc.} We
see then that the quantum data processing needed to extract a syndrome can be
carried out {\it at all levels of the concatenated code
simultaneously}.\footnote{A {\it destructive} measurement of the encoded
ancilla block can be carried out at all levels simultaneously.  The procedure
for measuring the block {\it nondestructively} (projecting the block onto
$|0\rangle_{\rm code}$ or $|1\rangle_{\rm code}$) is much more laborious; it
must be carried out on one level at a time.} After some relatively
straightforward classical processing, we determine what single qubit gates need
to be applied to all the elementary qubits in order to complete the recovery
step on all levels at once.

Thus it is easy to see (at least conceptually) how the accuracy threshold can
be estimated.\cite{gottesman_e}  At each level of the concatenated code, a
block of 7 fails if there are errors in at least two of the subblocks that it
contains.  If $p_L$ is the probability of an error in a block at level $L$,
then the probability of an error in a block at level $L+1$ is
\begin{equation}
\label{seven_choose}
p_{L+1}\sim \left({7\atop 2}\right) p_L^2 + \cdots= 21 ~p_L^2  +\cdots
\end{equation}
(neglecting the terms of higher order in $p_L$), which will be smaller than
$p_L$ if $p_L< 1/21$.  Therefore, if the each elementary qubit has a
probability of error $p_0 < 1/21$, the error probability will be smaller at
level 1, still smaller at level 2, and so on---the threshold value of $p_0$ is
$1/21$.

Suppose that we perform error correction every time we execute an XOR or single
qubit gate. Roughly speaking, $p_0$ is the probability of error per data qubit
when a cycle of error correction begins. To estimate the accuracy threshold, we
follow the circuit Fig.~\ref{fig:steane_comp} and add up the contributions to
$p_0$ due to errors (including possible storage errors) that arose during
recently executed quantum gates  and that have not already been eliminated in a
previous error correction cycle.  We obtain an expression for $p_0$ in terms of
the gate error and storage error probabilities that we can equate to $1/21$ to
find the threshold.

Proceeding this way, assuming that storage errors are negligible, and that each
single-qubit or XOR gate has the same error probability $\epsilon_{\rm gate}$,
we\cite{gottesman_e} crudely estimate the threshold gate error rate as
\begin{equation}
\label{thresh_gate}
\epsilon_{\rm gate,0}\sim 6\cdot 10^{-4} \ .
\end{equation}
Similarly, if gate errors are negligible, the estimated threshold storage error
rate is
\begin{equation}
\label{thresh_store}
\epsilon_{\rm store,0}\sim 6\cdot 10^{-4} \ .
\end{equation}
The thresholds for gate and storage errors are essentially the same because the
Steane method is well optimized for dealing with storage errors. The qubits are
rarely idle; a gate acts on each one in almost every step.  Hence, the storage
accuracy requirement is considerably less stringent than in previous threshold
estimates based on the Shor recovery
method.\cite{gottesman_c,gottesman_d,preskill_a}

However, a more thorough analysis shows that, for several reasons, the actual
threshold that can be inferred from the circuit Fig.~\ref{fig:steane_comp} is
somewhat lower than the estimates Eq.~(\ref{thresh_gate}) and
Eq.~(\ref{thresh_store}).   The most serious caveat is that to perform recovery
we must have a supply of well verified $|0\rangle_{\rm code}$ states encoded at
level $L$.  A separate (and rather complicated) calculation is required to
determine the threshold for reliable encoding.  We also need to analyze Shor's
implementation of the Toffoli gate to ensure that highly reliable Toffoli gates
can be executed on the concatenated blocks.\footnote{The elementary Toffoli
gates are not required to be as
accurate as the one and two-body gates -- an Toffoli gate error rate of order
$10^{-3}$ is acceptable, if the other error rates are sufficiently small. This
finding is welcome, since Tofolli gates are more difficult to implement, and
are likely to be less accurate in practice.} Finally, we must bound the
higher-order contributions to the failure probability that have been dropped in
Eq.~(\ref{seven_choose}) to obtain a rigorous result.  The full analysis for
this case has not yet been completed, but it seems conservative to guess that
the final values of the storage and gate thresholds will exceed $10^{-4}$.  Of
course, it is possible that with a better coding scheme and/or error recovery
protocol  a much higher value of the accuracy threshold could be established.

We should also ask how large a block size is needed to ensure a certain
specified accuracy. Roughly speaking, if the threshold gate error rate is
$\epsilon_0$ and the actual elementary gate error rate is $\epsilon<
\epsilon_0$, then concatenating the code $L$ times will reduce the error rate
to
\begin{equation}
\epsilon^{(L)}\sim  \epsilon_0\left({\epsilon\over\epsilon_0}\right)^{2^L} \ ;
\end{equation}
thus, to be reasonably confident that we can complete a computation with $T$
gates without making an error we must choose the block size $7^L$ to be 
\begin{equation}
\label{scaling}
{{\rm block}\atop{\rm size}} \sim \left[ {\log \epsilon_0 T}\over {\log
\epsilon_0/\epsilon}\right]^{\log_2 7} \quad .
\end{equation}
If the code that is concatenated has block size $n$ and can correct $t+1$
errors, the power  $\log_2 7\sim 2.8$ in Eq.~(\ref{scaling}) is replaced by
$\log
n / \log (t+1)$;  this power approaches 2 for the family of codes considered by
Shor, but could in principle approach 1  for ``good'' codes.

When the error rates are below the accuracy threshold, it is also possible to
maintain an {\it unknown} quantum state for an indefinitely long time.
However, as we have already noted in \S3.5, if the probability
of a storage error per computational time step is $\epsilon$, then the initial
encoding of the state can be performed with a fidelity no better than
$F=1-O(\epsilon)$.  With concatenated coding, we can store unknown quantum
information with reasonably good fidelity for an indefinitely long time, but we
cannot achieve arbitrarily good fidelity.

Concatenation is an important theoretical construct, since it enables us to
establish that arbitrarily long computations are possible.  But unless the
error rates are quite close to the threshold values, concatenated coding may
not be the best way to perform a particular computation of given length.
Indeed, a code chosen from the family originally described by Shor may turn out
to be more efficient that the concatenated 7-bit code.  Furthermore, the
concatenated 7-bit code {\it and} Shor's codes encode just a single qubit of
quantum information in a rather large code block.  But we saw in \S4 that
fault-tolerant quantum computation can be carried out using any stabilizer
code, including codes that make more efficient use of storage space by encoding
many qubits in a single block.  If the reliability of our hardware is close to
the accuracy threshold, then efficient codes will not work effectively.  But as
the hardware improves, we can use better codes, and so enhance the reliability
of our quantum computer at a smaller cost in storage space.

\section{Error models}
\label{dream}
A fault-tolerant scheme should be tailored to protect against the types of
errors that are most likely to afflict a particular device.  And any statement
about what error rates are acceptable (like the estimate of the accuracy
threshold that we have just outlined) is meaningless unless a model for the
errors is carefully specified.
Let's summarize some of the important assumptions about the error model that
underlie our estimate of
the accuracy threshold:

\begin{itemize}
\item {\bf Random errors.} We have assumed that the errors have no systematic
component.\footnote{Knill {\it et al.}~\cite{knill_c} have demonstrated the
existence of an accuracy threshold for much more general error models.}  Errors
that have random phases accumulate like a random walk, so that the {\it
probability} of error accumulates roughly linearly with the number of gates
applied.  But it the errors have systematic phases, then the error {\it
amplitude} can increase linearly with the number of gates applied. Hence, for
our quantum computer to perform well, the rate for systematic errors must meet
a more stringent requirement than the rate for random errors.  Crudely
speaking, {\it if} we assume that the systematic phases always conspire to add
constructively, and if the accuracy threshold is
$\epsilon_0$ for the case of random errors, then the accuracy threshold would
be of order $(\epsilon_0)^2$ for
(maximally conspiratorial) systematic errors.  While systematic errors may pose
a challenge to the quantum engineers of the future, they ought not to pose an
insuperable obstacle. My attitude is that (1) even if
our hardware is susceptible to making errors with systematic phases, these will
tend to cancel out in the course of a reasonably long
computation,~\cite{obenland_a}$^{\!-\,}$\cite{miquel} and (2) since systematic
errors can
in principle be understood and eliminated, from a fundamental point of view it
is more relevant to know the limitations on the performance of the machine that
are imposed by the random errors. 

\item {\bf Uncorrelated errors.} We have assumed that the errors are both
spatially and temporally uncorrelated with one another.  Thus when we say that
the probability of error per qubit is (for example) $\epsilon\sim 10^{-5}$, we
actually mean
that, given two specified qubits, the probability that errors afflict both is
$\epsilon^2\sim 10^{-10}$.  This is a very strong assumption.  The really
crucial requirement is that correlated errors affecting multiple qubits in the
same code block are highly unlikely,
since our coding schemes will fail if several errors occur in a single block.
Future quantum engineers will face the challenge of designing devices such that
qubits in the same block are very well isolated from one another. 

\item {\bf Maximal parallelism.} We have assumed that many quantum gates can be
executed in parallel in a single time step.  This assumption enables us to
perform error recovery in all of our code blocks at once, and so is critical
for controlling qubit storage errors.  (Otherwise, if we added a level of
concatenation to the code, each individual resting qubit would have to wait
longer for us to get around to perform recovery, and would be more likely to
fail.) If we ignore storage errors, then parallel operation is not essential in
the analysis of the accuracy threshold, but it would certainly be desirable to
speed up the computation.

\item {\bf Error rate independent of number of qubits.}  We have assumed that
the error rates do not depend on how many qubits are stored in our device.
Implicitly, this is an assumption about the nature of the hardware.  For
example, it would not be a reasonable assumption if all of the qubits were
stored in a single ion trap, and all shared the same phonon bus.~\cite{cirac}

\item {\bf Gates can act on any pair of qubits.}  We have assumed that our
machine is equipped with a set of fundamental gates that can be applied to any
pair of stored qubits (or triplet of qubits, in the case of the Toffoli gate),
irrespective of their proximity.  In practice, there is likely to be a cost,
both in processing time and error rate, of shuttling the qubits around so that
a gate can act effectively on a particular pair.  We leave it to the machine
designer to choose an architecture that minimizes this cost.  If gates can act
only on neighboring qubits, there will still be a threshold,\cite{aharonov} but
it will be considerably lower.

\item {\bf Fresh ancilla qubits.} We have assumed that our computer has access
to an adequate supply of fresh ancillary qubits.  The ancilla qubits are used
both to implement (Toffoli) gates and to perform error recovery.  As the
effects of random errors accumulate, entropy is generated, and the error
recovery process flushes the entropy from the computing device into the ancilla
registers.  In principle, the computation can proceed indefinitely as long as
fresh ancilla qubits are provided, but in practice we will want to clear the
ancilla and reuse it.  Erasing the ancilla will necessarily dissipate power and
generate heat; thus cooling of the device will be required.

\item {\bf No leakage errors.}  We have ignored the possibility of {\it
leakage}.  In
our model of a quantum computer, each of our qubits lives in a two-dimensional
Hilbert space, and we assumed that, when an error occurs, this qubit either
becomes entangled with the environment or rotates in the two-dimensional space
in an unpredictable way.  But there is another possible type of error, in which
the qubit leaks out of the two-dimensional space into a larger
space.\cite{plenio}  To control leakage errors, we can repeatedly interrogate
each qubit to test for leakage (for example, using the leakage-detection
circuit shown in Fig.~\ref{fig_leak}), without trying to diagnose exactly what
happened to the leaked qubit.\cite{preskill_a} If leakage has occurred, the
qubit is damaged and must be
discarded;~\footnote{Of course, we can recycle it later.}  we replace it with a
fresh qubit in a standard state, say the
state $|0\rangle$. Then we can perform conventional
syndrome measurement, which will project the qubit onto a state such that the
error can be reversed by a simple unitary transformation.\footnote{In fact,
since we
know before the syndrome measurement that the damaged qubit is in a particular
position within the code block, we can apply a streamlined version of error
correction designed to diagnose and reverse the error at that known
position.~\cite{grassl}}
If concatenated coding is used, leakage detection need be
implemented only at the lowest coding level. The detection circuit is quite
simple, so allowing leakage errors does not have much effect on the accuracy
threshold.

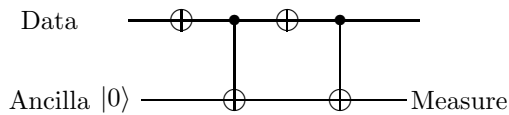
\begin{figure}
\centering
\begin{picture}(180,50)

\put(25,10){\makebox(0,0){$|0\rangle$}}
\put(0,10){\makebox(0,0){Ancilla}}
\put(0,40){\makebox(0,0){Data}}

\put(35,10){\line(1,0){100}}
\put(30,40){\line(1,0){110}}

\put(50,44){\line(0,-1){8}}
\put(50,40){\circle{8}}

\put(70,40){\circle*{4}}
\put(70,40){\line(0,-1){34}}
\put(70,10){\circle{8}}

\put(90,44){\line(0,-1){8}}
\put(90,40){\circle{8}}

\put(110,40){\circle*{4}}
\put(110,40){\line(0,-1){34}}
\put(110,10){\circle{8}}

\put(155,10){\makebox(0,0){Measure}}

\end{picture}
\caption{A quantum leak detection circuit.  Assuming that the XOR gate acts
trivially if the data has leaked, then the outcome of the measurement is 0
if leakage has occurred, 1 otherwise.}
\label{fig_leak}
\end{figure}

\end{itemize}

The assumptions of our error model are sufficiently realistic to provide
reasonable guidance concerning how well a quantum computer can perform under
noisy conditions. Suppose, for example, that we want our quantum computer to
solve a hard factoring problem using Shor's algorithm; what specifications must
be met by the machine?  With the best known classical factoring algorithm and
the fastest existing machines, it takes a few months to factor a 130 digit
(432-bit) number.~\cite{lenstra} To perform this task with Shor's algorithm, we
would need to be able to store about $5\cdot 432 = 2160$ qubits and to perform
about $38\cdot (432)^3\sim 3\cdot 10^9$ Toffoli gates.~\cite{beckman}. To have
a reasonable chance of performing the computation with acceptable accuracy, we
would want the probability of error per Toffoli gate to be less than about
$10^{-9}$, and the probability of a storage error per gate execution time to be
less than about $10^{-12}$.

According to the concatenation flow equations for the 7-qubit
code,\footnote{This analysis\cite{preskill_a}
 was actually carried out for the Shor method of syndrome measurement, rather
than the Steane method invoked in our discussion of concatenated coding in
\S5.}these error rates can be achieved for the encoded data, if the error rates
at
the level of individual qubits are $\epsilon_{\rm store}\sim \epsilon_{\rm
gate}\sim 10^{-6}$, and if 3 levels of concatenation are used, so that the size
of the block encoding each qubit is $7^3=343$.  Allowing for the additional
ancilla qubits needed to implement gates and (parallelized) error correction,
the total number of qubits required in the machine would be of order $10^6$.

When the storage error rate is fairly high, concatenation may be the most
effective coding procedure.  But if gate errors dominate (and if the gate error
rate is not too close to the threshold), then other quantum codes give a better
performance.   For example, Steane~\cite{steane_d} found that this same
factoring problem could be solved by a quantum computer with $4\cdot 10^5$
qubits and a gate error rate of order $10^{-5}$, using a code with block size
55 that can correct 5 errors.  At lower error rates it is
possible to use codes that make more efficient use of storage space by encoding
many qubits in a single block.\cite{gottesman_b}  

Surely, a quantum computer with about a million qubits and an error
rate per gate of about one in a million would be a very powerful and valuable
device
(assuming a reasonable processing speed). Of course, from the perspective of
the current
state of the technology,\cite{monroe}$^{\!-\,}$\cite{gershenfeld} these numbers
seem
daunting.  But in fact a machine that meets far less demanding specifications
may still be very useful.\cite{preskill_b} First of all, quantum computers can
do other things besides factoring, and some of these other tasks (in particular
quantum simulation\cite{lloyd_a}) might be accomplished with a less reliable
or smaller device.  Furthermore, our estimate of the accuracy threshold might
be too conservative for a number of reasons. For example, the estimate was
obtained under the assumption that phase and amplitude errors in the qubits are
equally likely.  With a more realistic error model better representing the
error probabilities in an actual device, the error correction scheme could be
better tailored to the error model, and a higher error rate could be tolerated.
Also, even under the assumptions stated, the fault-tolerant scheme has not
been definitively analyzed; with a more refined analysis, one can expect to
find a somewhat higher accuracy threshold, perhaps considerably higher.
Substantial improvements might also be attained by modifying the fault-tolerant
scheme, either by finding a more efficient way to implement a universal set of
fault-tolerant gates, or by finding a more efficient means of carrying out the
measurement of the error syndrome.  With various improvements, it would not be
surprising\footnote{In fact,
estimates of the accuracy threshold that are more optimistic than mine have
been put forward by Zalka.\cite{zalka}}  to find that a quantum computer could
work effectively with a
probability of error per gate, say, of order $10^{-3}$. 

An error rate of, say $10^{-5}$ is surely ambitious,
but not, perhaps, beyond the scope of what might be achievable in the future.
In any case, we now have a fair notion of how good the performance of a useful
quantum computer will need to be.  And that in itself represents enormous
progress over just two years ago.

\section{Topological Quantum Computation}
\subsection{Aharonov-Bohm Phenomena and Superselection Rules}
Now that we know that quantum error correction is possible, it is important to
broaden our perspective --- we should strive to go beyond the analysis of
abstract circuits and explore the potential physical contexts in which quantum
information might be reliably stored and manipulated.  In particular, we might
hope to design quantum gates that are {\it intrinsically} fault tolerant, so
that active intervention by the computer operator will not be required to
protect the machine from noise. A significant step toward this goal has been
taken recently by Alexei Kitaev;\cite{kitaev_c} this section is based on his
ideas.

Topological concepts have a natural place in the discussion of quantum error
correction and fault-tolerant computation.  Topology concerns the ``global''
properties of an object that remain unchanged when we deform the object
locally.  The central idea of quantum error correction is to store and
manipulate quantum information in a ``global'' form that is resistant to local
disturbances.   A fault-tolerant gate should be designed to act on this global
information, so that the action it performs on the encoded data remains
unchanged even if we deform the gate slightly; that is, even if the
implementation of the gate is not perfect.

In seeking physical implementations of fault-tolerant quantum computation,
then, we  ask whether there are known systems in which physical interactions
have a topological character.  Indeed, topology is at the essence of the {\it
Aharonov-Bohm effect}.  If an electron is transported around a perfectly
shielded magnetic solenoid, its wave function acquires a phase $e^{ie\Phi}$,
where $e$ is the electron charge and $\Phi$ is the magnetic flux enclosed by
the solenoid.  This Aharonov-Bohm phase is a topological property of the path
traversed by the electron --- it depends only on how many times the electron
circumnavigates the solenoid, and is unchanged when the path is smoothly
deformed.  (See Fig.~\ref{fig:ab}.) We are thus led to contemplate a
realization of quantum computation in which information is encoded in a form
that can be measured and manipulated through Aharonov-Bohm interactions ---
topological interactions that are immune to local disturbances.

\begin{figure}
\hskip 1in
\psfig{figure=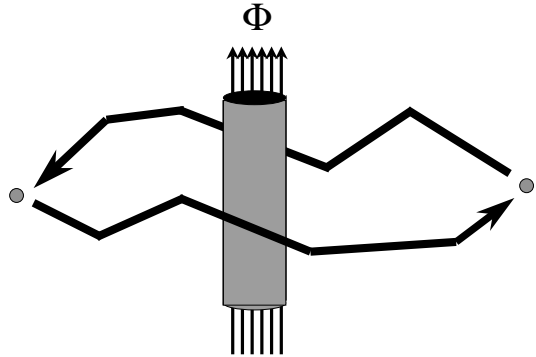,height=2.5in}
\caption{A topological interaction.  The Aharonov-Bohm phase acquired by an
electron that encircles a flux tube remains unchanged if the electron's path is
slightly deformed.\label{fig:ab}}
\end{figure}

It is useful to reexpress this reasoning in the language of superselection
rules.  A superselection rule, as I am using the term here, arises (in a field
theory or spin system defined in an infinite spatial volume) if Hilbert space
decomposes into mutually orthogonal sectors, where each sector is preserved by
any local operation.  Perhaps the most familiar example is the charge
superselection rule in quantum electrodynamics.  An electric charge has an
infinite range electric field. Therefore no local action can create or destroy
a charge, for to destroy a charge we must also destroy the electric field lines
extending to infinity, and no local procedure can accomplish this task.

The Aharonov-Bohm interaction is also an infinite range effect; the electron
acquires an Aharonov-Bohm phase upon circling the solenoid no matter what its
distance from the solenoid. So we may infer that no local operation can destroy
a charge that participates in Aharonov-Bohm phenomena.  If we consider two
objects carrying such charges, widely separated and well isolated from other
charged objects, then any process that changes the charge on either object
would have to act coherently in the whole region containing the two charges.
Thus, the charges are quite robust in the presence of localized disturbances;
we can strike the particle with a hammer or otherwise abuse it without
modifying the charges that it carries.

Following Kitaev,\cite{kitaev_c} we may envision a {\it topological quantum
computer}, a device in which quantum information is encoded in the quantum
numbers carried by quasiparticles that reside on a two--dimensional surface and
have long-range Aharonov-Bohm interactions with one another. At zero
temperature, an accidental exchange of quantum numbers between quasiparticles
(an error) arises only due to quantum tunneling phenomena involving the virtual
exchange of charged objects.  The amplitude for such processes is of the order
of $e^{-m L}$,  where $m$ is the mass of the lightest charged object (in
natural units), and $L$ is the distance between the two quasiparticles.  If the
quasiparticles are kept far apart, the probability of an error afflicting the
encoded information will be extremely low. At finite temperature $T$, there is
an additional source of error, because an uncontrolled plasma of charged
particles will inevitably be present, with a density proportional to the
Boltzman factor $e^{-\Delta/T}$, where $\Delta$  is the mass gap (not
necessarily equal to the ``curvature mass'' $m$). Sometimes one of the plasma
particles will slip unnoticed between two of our data-carrying particles,
resulting in an exchange of charge and hence an error. To achieve an acceptably
low error rate, then, we would need to keep the temperature well below the gap
$\Delta$ (or else we would have to monitor the thermal plasma very faithfully).

\subsection{The Fractional Quantum Hall Effect and Beyond}

If our device is to be capable of performing interesting computations, the
Aharonov-Bohm phenomena that it employs must be {\it nonabelian}.  Only then
will we be able to build up complex unitary transformations by performing many
particle exchanges in succession. Such nonabelian Aharonov-Bohm effects can
arise in systems with nonabelian gauge fields.  Nature has been kind enough to
provide us with some fundamental nonabelian gauge fields, but unfortunately not
very many, and none of these seem to be suited for practical quantum
computation.  To realize Kitaev's vision, then, we must hope that nonabelian
Aharonov-Bohm effects can arise as complex collective phenomena in
(two-dimensional electron or spin) systems that have only short-range
fundamental interactions.

In fact, one of the most remarkable discoveries of recent decades has been that
infinite range Aharonov-Bohm phenomena {\it can} arise in such systems, as
revealed by the observation of the fractional quantum Hall effect.  The
electrons in quantum Hall systems are so highly frustrated that the ground
state is an extremely entangled state with strong quantum correlations
extending out over large distances. Hence, when one quasiparticle is
transported around another, even when the quasiparticles are widely separated,
the many electron wave function acquires a nontrivial Berry phase (such as
$e^{2\pi i/3}$).  This Berry phase is indistinguishable in all its observable
effects from an Aharonov-Bohm phase arising from a fundamental gauge field, and
its experimental consequences are spectacular.\cite{fqhe}

The Berry phases observed in quantum Hall systems are abelian (although there
are some strong indications that nonabelian Berry phases can occur under the
right conditions\cite{read,wilczek}), and so are not very interesting from the
viewpoint of quantum computation.  But Kitaev\cite{kitaev_c} has described a
family of simple spin systems with local interactions in which the existence of
quasiparticles with nonabelian Berry phases can be demonstrated.  (The
Hamiltonian of the system so frustrates the spins that the ground state is a
highly entangled state with infinite range quantum correlations.) These models
are sufficiently simple (although unfortunately they require four-body
interactions), that one can imagine a designer material that can be reasonably
well-described by one of Kitaev's models.  The crucial topological properties
of the model are relatively insensitive to the precise microscopic details, so
the task of the fabricator who ``trims'' the material may not be overly
demanding.  If furthermore it were possible to control the transport of
individual quasiparticles (perhaps with a suitable magnetic tweezers), then the
system could be operated as a fault-tolerant quantum computer.

\begin{figure}
\hskip 1in
\psfig{figure=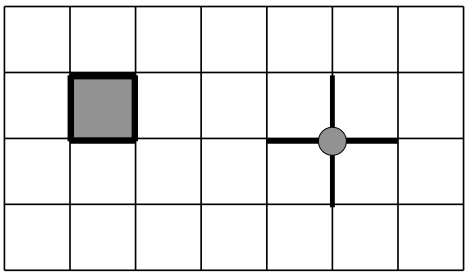,height=2.5in}
\caption{A Kitaev spin model. Spins reside on the lattice links.  The four
spins that meet at a site or share a plaquette are coupled.\label{fig:lattice}}
\end{figure}

To construct his models, Kitaev considers a square lattice, with spins residing
on each lattice link.  The Hamiltonian is expressed as a sum of mutually
commuting four-body operators, one for each site and one for each plaquette of
the lattice.  (See Fig.~\ref{fig:lattice}.)  Because the terms are mutually
commuting, it is simple to diagonalize the Hamiltonian by diagonalizing each
term separately.  The operators on sites resemble local gauge symmetries
(acting independently at each site), and a state that minimizes these terms is
invariant under the local symmetry, like the physical states that obey Gauss's
law in a gauge theory. The operators on plaquettes are like ``magnetic flux''
operators in a gauge theory, and these terms are minimized when the magnetic
flux vanishes everywhere.  The excitation spectrum includes states in which
Gauss's law is violated at isolated sites --- these points are ``electrically
charged'' quasiparticles ---  and states in which the magnetic flux is
nonvanishing at isolated plaquettes --- these are magnetic fluxon
quasiparticles. The quantum entanglement of the ground state is such that a
nontrivial Berry phase is associated with the transport of a charge around a
flux --- this phase is identical to the Aharonov-Bohm phase in the analog gauge
theory.

These Aharonov-Bohm phenomena are stable even as we deform the Hamiltonian of
the theory.  Indeed, if the deformation is sufficiently small, we can study its
effects using perturbation theory.  But as long as the perturbations are local
in space, topological effects are robust, since perturbation theory is just a
sum over localized influences.  Whatever destroys the long-range topological
interactions must be nonperturbative in the deformation of the theory.

Two types of nonperturbative effects can be anticipated.\cite{thooft}  The
ground state of the theory might become a ``flux condensate'' with an
indefinite number of magnetic excitations.  In this event, there would be a
long-range attractive interaction between charged particles and their
antiparticles.  It would be impossible to separate charges, and there would be
no long-range effects.  In a gauge theory, this phenomenon would be called {\it
electric confinement}.  Alternatively, a condensate of electric quasiparticles
might appear in the ground state.  Then the magnetic excitations would be
confined, and again the long-range Aharonov-Bohm effects would be destroyed.
In a gauge theory, we would call this the Higgs phenomenon (or magnetic
confinement).

Thus, as we deform Kitaev's Hamiltonian, we can anticipate that a phase
boundary will eventually be encountered, beyond which either electric
confinement or the Higgs phenomenon will occur.  The size of the region
enclosed by this boundary will determine how precisely a material will need to
be fabricated in order to behave as Kitaev specifies.  A particularly urgent
question for the material designer is whether cleverly chosen {\it two-body}
interactions might so frustrate a spin system as to produce a highly entangled
ground state and nonabelian Aharonov-Bohm interactions among the quasiparticle
excitations.

The fractional quantum Hall effect, and Kitaev's models, speak a memorable
lesson.  We find gauge phenomena emerging as collective effects in systems with
only short range interactions.  It is intriguing to speculate that the gauge
symmetries known in Nature could have a similar origin.

\subsection{Topological Interactions}
As we have noted, in Kitaev's spin models, there are two types of charges that
can be carried by localized quasiparticles, which we may call ``electric'' and
``magnetic'' charges. In the simplest type of model, the ``magnetic flux''
carried by a particle can be labeled by an element of a finite group $G$, and
``electric charges'' are labeled by irreducible representations\footnote{There
can also be ``dyons'' that carry both types of charge, and the classification
of the charge carried by a dyon is somewhat subtle, but we will not need to
discuss explicitly the properties of the dyons.} of $G$.  If a charged particle
in the irreducible representation $D^{(\nu)}$, whose quantum numbers are
encoded in an internal wavefunction $|\psi^{(\nu)}\rangle$,  is carried around
a flux labeled by group element $u\in G$, then the wavefunction is modified
according to
\begin{equation}
|\psi^{(\nu)}\rangle \to D^{(\nu)}(u)|\psi^{(\nu)}\rangle \ .
\end{equation}
Exploiting this interaction, we can {\it measure} a magnetic flux by scattering
a suitable charged particle off of the flux.\cite{alford_a}  For example, we
could construct a Mach-Zender flux interferometer as shown in
Fig.~\ref{fig:inter} that is sensitive to the relative phase acquired by the
charged particle paths that pass to the left or right of the flux.  If we
balance the interferometer properly, we can distinguish between, say,  two flux
values $u_1,u_2\in G$; a $u_1$ flux will be detected emerging from one arm of
the interferometer, and a $u_2$ flux from the other arm.  Of course, the
interferometer we build will not be flawless, but the flux measurement can
nevertheless be fault-tolerant --- if we have many charged projectiles and
perform the measurement repeatedly, we can determine the flux with very high
statistical confidence.

\begin{figure}
\hskip 1in
\psfig{figure=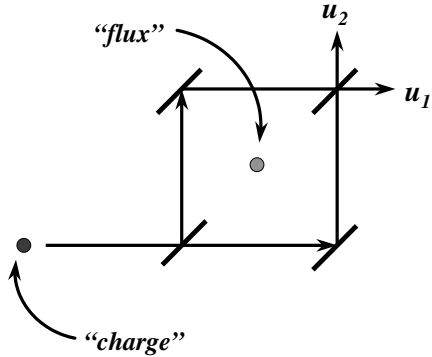,height=2.5in}
\caption{A Mach-Zender interferometer for flux measurement, shown
schematically. The flux to be measured is inserted inside.  The test charge
emerges from one arm if the flux has value $u_1$, the other arm if the flux has
value $u_2$. \label{fig:inter}}
\end{figure}

If the two fluxes $u_1$ and $u_2$ belong to the same conjugacy class in $G$,
then there is a symmetry relating the two fluxons, so that all local physics is
indifferent to the value of the flux (see below).  Therefore, a coherent
superposition of fluxes
\begin{equation}
a|u_1\rangle + b |u_2\rangle
\end{equation}
will not readily decohere due to localized interactions with the environment.
But the flux interferometer (operated repeatedly) will project the fluxon onto
either of the flux eigenstates $|u_1\rangle$ (with probability $|a|^2$) or
$|u_2\rangle$ (with probability $|b|^2$).

\begin{figure}
\hskip 1in
\psfig{figure=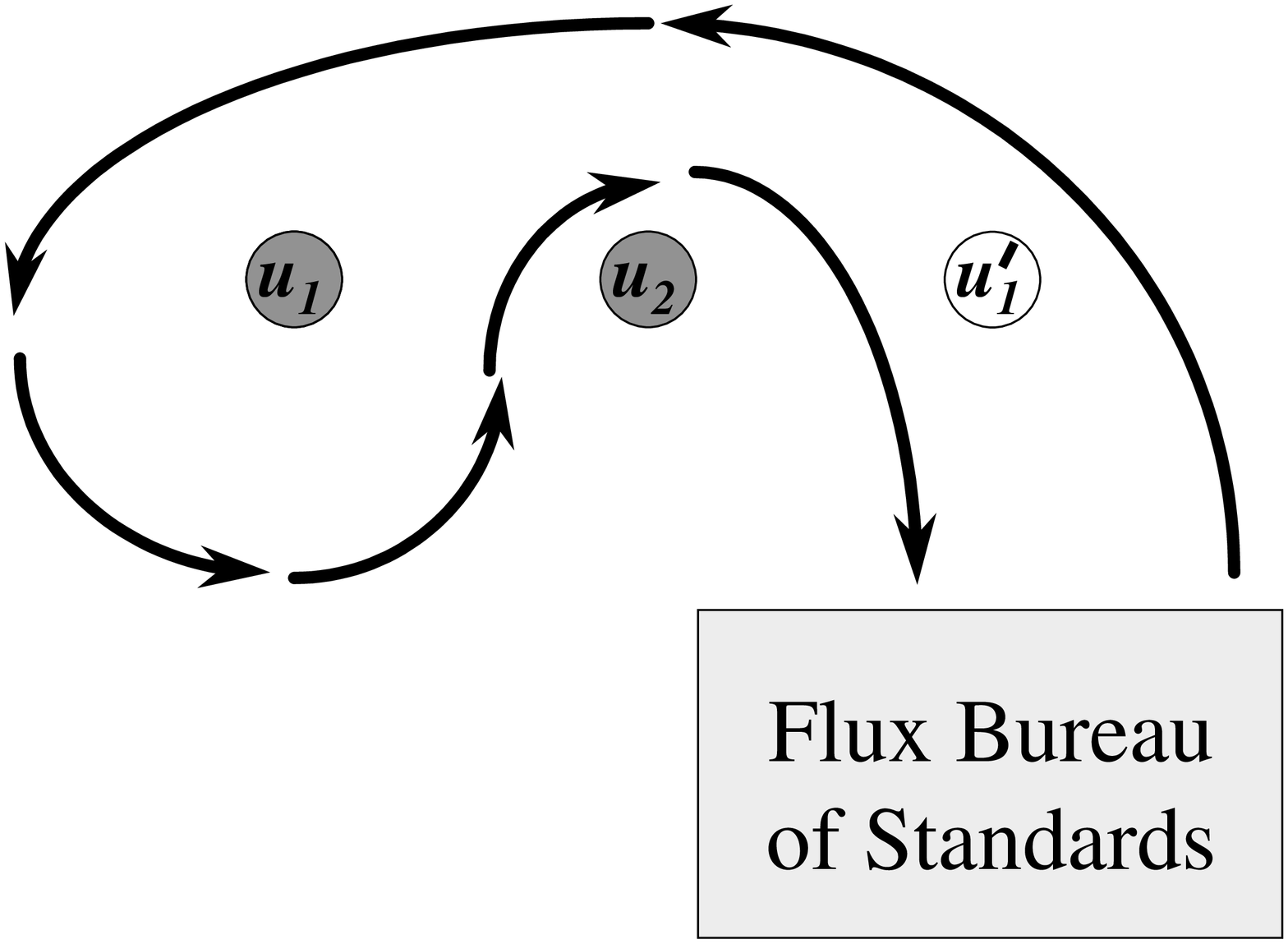,height=2.5in}
\caption{The flux exchange interaction.  The flux labeled $u_1$ is carried from
its original position (shaded) to its new position (unshaded), and then
remeasured.  The charged particle path shown that encircles the original
position of the flux  is topologically equivalent to a path that encircles the
new position; hence the value of the flux changes from $u_1$ to
$u_1'=u_2^{-1}u_1 u_2$. \label{fig:exchange}}
\end{figure}

Now imagine that two fluxons have been carefully calibrated, so that one is
known to carry the flux $u_1$ and the other the flux $u_2$.  And suppose that
the two vortices are carefully ``exchanged'' by carrying the first around the
second as shown in Fig.~\ref{fig:exchange}, and that we subsequently remeasure
the fluxes.  Carrying a charged particle around the fluxon on the right, after
the exchange, is topologically equivalent to carrying the charged particle
around first the right fluxon, then the left fluxon, and finally the right
fluxon in the opposite direction, before the exchange.  We infer that the
exchange modifies the quantum numbers of the fluxons according to
\begin{equation}
\label{flux_braid}
|u_1\rangle|u_2\rangle \to |u_2\rangle|u_2^{-1}u_1 u_2\rangle \ ,
\end{equation}
a nontrivial  interaction if the two fluxes fail to commute.\cite{bais}  Thus,
noncommuting fluxes have interesting Aharonov-Bohm interactions of their own,
even in the absence of any electric charges. Because carrying one flux around
another can {\it conjugate} the value of the flux, two fluxons carrying
conjugate fluxes must be regarded as {\it indistinguishable}
particles.\cite{lo}  An exchange of two such objects can modify their internal
quantum numbers; we will refer to them as {\it nonabelions},\cite{moore}
indistinguishable particles in two dimensions that obey an exotic nonabelian
variant of quantum statistics.

\begin{figure}
\hskip 1in
\psfig{figure=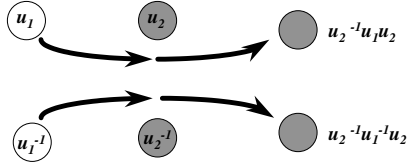,height=2.5in}
\caption{The ``pull-through'' interaction.  One flux pair is pulled through
another.  The outside flux is unmodified, but the inside flux is conjugated by
the outside flux. \label{fig:pull}}
\end{figure}

We will use the exchange interaction Eq.~(\ref{flux_braid}) as a fundamental
logical operation in our Aharonov-Bohm quantum computer.  However, it will
actually be convenient to encode qubits in pairs of fluxons, where the total
flux of the pair is trivial.\cite{kitaev_c}  That is, we will consider
fluxon-antifluxon pairs of the form $|u,u^{-1}\rangle$, but where the flux and
antiflux are kept far enough apart from one another that an inadvertent
exchange of quantum numbers between them is unlikely.  To perform logic, we may
pull one pair through another as shown in Fig.~\ref{fig:pull}.  Since the total
flux that passes through the middle of the outside pair is trivial, this pair
is not modified, but the inside fluxes are conjugated by the outside flux:
\begin{equation}
\label{pair_braid}
|u_1,u_1^{-1}\rangle|u_2, u_2^{-1}\rangle \to |u_2,u_2^{-1}\rangle|u_2^{-1}u_1
u_2, u_2^{-1} u_1^{-1} u_2\rangle \ ;
\end{equation}
an operation that is evidently isomorphic to the effect of the exchange of
single fluxes described by Eq.~(\ref{flux_braid}).  Using pairs instead of
single fluxons has two advantages.  First, since each pair has trivial total
flux, the pairs do not interact unless one is pulled through another;
therefore, we can easily shunt pairs around the device without inducing any
unwanted interactions with distant pairs.  Second, and more important, pairs
can carry charges even if each member of the pair carries no
charge.\cite{alford_b,preskill}  The charge of a pair can be measured, and this
charge-measurement operation will be a crucial ingredient in the construction
of a universal set of quantum gates.  The operation Eq.~(\ref{pair_braid}) can
be regarded as a {\it classical} logic gate; it takes flux eigenstates to flux
eigenstates.  To perform interesting quantum computations, we will need to be
able to prepare coherent superpositions of flux eigenstates.  This is what we
can accomplish by measuring the charge of a pair.

\begin{figure}
\hskip 1in
\psfig{figure=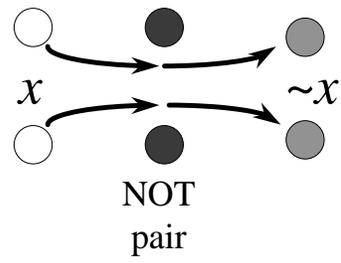,height=2.5in}
\caption{The NOT gate. Pulling a computational flux pair through a NOT pair
flips the value of the encoded bit.\label{fig:gate}}
\end{figure}

\begin{figure}
\hskip 1in
\psfig{figure=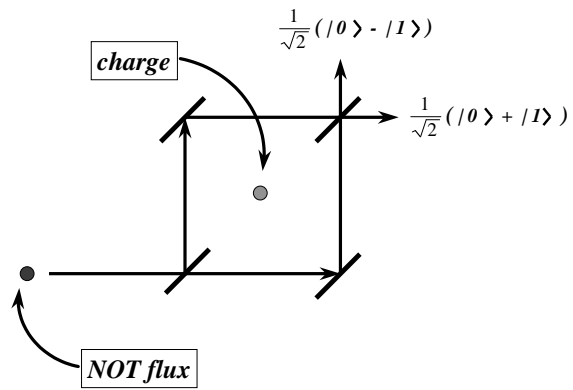,height=2.5in}
\caption{A Mach-Zender interferometer for charge measurement, shown
schematically.  The flux pair whose charge is to be measured is inserted
inside.  If the test NOT flux emerges from one arm, the $|+\rangle$ charge
state has been prepared; if it emerges from the other arm, $|-\rangle$ has been
prepared. \label{fig:charge}}
\end{figure}

Suppose that $u_0$ and $u_1\in G$ are related by $u_1=v^{-1}u_0v$ for some
$v\in G$.  Then if we think of the flux eigenstates $|u_0, u_0^{-1}\rangle$ and
$|u_1,u_1^{-1}\rangle$ as computational basis states, the effect of pulling
either pair through a  $|v,v^{-1}\rangle$ pair can be interpreted as a NOT or
$X$ gate:
\begin{equation}
|u_0, u_0^{-1}\rangle \leftrightarrow |u_1,u_1^{-1}\rangle
\end{equation}
(see Fig.~\ref{fig:gate}).
But suppose we wish to prepare one of the states
\begin{equation}
|\pm\rangle= {1\over \sqrt{2}}\left(|u_0, u_0^{-1}\rangle \pm
|u_1,u_1^{-1}\rangle\right) \ .
\end{equation}
We can project a coherent superposition of $|u_0, u_0^{-1}\rangle$ and
$|u_1,u_1^{-1}\rangle$ onto the $\{|\pm\rangle\}$ basis by scattering a
$|v\rangle$ fluxon off the pair, or in other words by operating a {\it charge
interferometer}, as in Fig.~\ref{fig:charge}. When the $|v\rangle$ fluxon
navigates around the pair, it acquires a trivial Aharonov-Bohm phase if the
pair is in the state $|+\rangle$ and the nontrivial phase $-1$ if the pair is
in the state $|-\rangle$.  If the interferometer is properly balanced, then,
the $|v\rangle$ projectile will be detected  emerging from one arm of the
interferometer if the pair is $|+\rangle$, and the other arm if the pair is
$|-\rangle$.  This is an example of charge measurement.  Though the
interferometer will not be perfect, charge measurement (like flux measurement)
can be fault-tolerant, if we repeat the measurement enough times.

\subsection{Universal Topological Computation}

Working with fluxon pairs as computational basis states, we have seen how to
perform the exchange (or ``pull through'') operation Eq.~(\ref{pair_braid}),
how to measure flux (using previously calibrated charges), and how to measure
charge (using previously calibrated fluxes).  We will also suppose that we are
able to produce a large supply of vortex pairs.  Local processes produce pairs
that carry no charge or flux; a charge-zero pair with trivial flux has the form
(up to normalization)
\begin{equation}
|{\rm charge~~zero}\rangle =\sum_u |u,u^{-1}\rangle \ ,
\end{equation}
where the sum ranges over a complete conjugacy class of $G$.  Because this
state is left invariant when conjugated by any element of $G$, it has trivial
Aharonov-Bohm interactions with any flux, and so carries no detectable charge.
After producing such a pair, we can perform flux measurement to project out one
of the flux eigenstate pairs $|u,u^{-1}\rangle$.  Performing many such
measurements on many pairs, we can assemble a large reservoir of calibrated
flux pairs that can be withdrawn as needed during the course of a computation.

But is our quantum computer universal --- can we closely approximate any
desired unitary transformation?  To address this issue, we recall the result
mentioned in \S4.2:  Universal {\it classical} computation, together with the
ability to perform the single-qubit gates $X$ and $Z$, and the ability to {\it
measure} $X$, $Y$, and $Z$, suffice for universal quantum
computation.\cite{gottesman_b}  In fact, there are groups $G$ such that the
operation Eq.~(\ref{pair_braid}) is sufficient for universal classical
computation.  We have found\cite{ogburn} that a Toffoli gate can be constructed
from Eq.~(\ref{pair_braid}) if $G=A_5$, the group of even permutations on five
objects.  We may, for example, choose computational basis states with
\begin{equation}
u_0=(125) \ , \quad u_1=(234) \ ;
\end{equation}
that is, we choose our computational fluxes to be three-cycles with one object
in common.   Then a Toffoli gate can be constructed from a total of 16
elementary ``pull-through'' operations; six ancilla pairs are also used to
catalyze this reaction.  No Toffoli gate was found in any group smaller than
$A_5$.\footnote{Kitaev had reported earlier that universal classical
computation is possible for $G=S_5$.}  Since $A_5$ is also the smallest of the
finite nonsolvable groups, it is tempting to conjecture that nonsolvablility is
a necessary condition for universal classical computation generated by
conjugation.\footnote{A finite group is {\it nonsolvable} if it has a
nontrivial subgroup whose commutator subgroup is itself.
Barrington\cite{barrington} also found evidence for a separation in the
computational complexity of group multiplication for solvable vs. nonsolvable
groups.}

We have already remarked that an $X$ gate can be realized by pulling a
computational vortex pair through the pair with flux $v$ such that
$u_1=v^{-1}u_0 v$; here we choose $v=(14)(35)$.  It turns out that the $Z$ gate
can be constructed with six pull-through steps and four ancilla pairs.
Measuring $Z$ is the same as measuring flux, and we have already seen that $X$
measurement can be achieved by measuring the charge of a pair, specifically, by
using a $v$ projectile in a charge interferometer.  It only remains to verify
that we can measure $Y$.  Though $Y$ measurement cannot be carried out exactly
in this scheme, it turns out that a {\it controlled}-$Y$ gate can be
constructed from 31 pull-through steps, and using 7 ancilla pairs.  Appealing
to another trick invented by Kitaev,\cite{kitaev_abelian} we can use the
controlled-$Y$ gate repeatedly to carry out $Y$-measurement to any desired
accuracy.\footnote{Actually, measuring $Y$ (which has eigenvalues $\pm i$)
using the controlled-$Y$ gate does not work, because the Kitaev method does not
distinguish between eigenvalues related by complex conjugation.  What we really
construct is a controlled-$\omega Y$ gate where $\omega=e^{2\pi i/3}$.}
Therefore, we have constructed a universal gate set using only the
Aharonov-Bohm interactions of fluxes and charges; we have a fault-tolerant
universal quantum computer.

Unfortunately, the spin model on which this construction is based is not so
simple.  Since the group $A_5$ has order 60, the Kitaev spin model that
realizes this scenario has a 60-component spin residing at each lattice link
(!)  One hopes that a simpler implementation of universal Aharonov-Bohm
computation will be found.

\subsection{Is Nature Fault Tolerant?}
The discovery of quantum error correction and fault tolerance has so altered
our thinking about quantum information that it is appropriate to wonder about
the potential implications for fundamental physics.  And in fact, a fundamental
issue pertaining to loss of quantum information has puzzled the physics
community for over twenty years.

In 1975, Stephen Hawking\cite{hawking} argued that quantum information is
unavoidably lost when a black hole forms and then subsequently evaporates
completely. The essence of the argument is very simple:  because of the highly
distorted causal structure of the black hole spacetime, the emitted radiation
is actually on the {\it same} time slice as the collapsing body that
disappeared behind the event horizon.  If the quantum information that is
initially encoded in the collapsing body is eventually to re-emerge encoded in
the microstate of the emitted information, then that information must be in two
places at once. In other words, the quantum information must be {\it cloned}, a
known impossibility under the usual assumptions of quantum
theory.\cite{dieks,wootters}  Hawking concludes that not all physical processes
can be governed by unitary time evolution; the laws of quantum theory need
revision.

This argument is persuasive, but many physicists are very distrustful of the
conclusion.  Perhaps one reason for the skepticism is that it seems odd for
Nature to tolerate just a little bit of information loss.\cite{banks} If
processes involving black holes can destroy information, then one expects that
information loss is unsuppressed at the Planck length scale
$(G\hbar/c^3)^{1/2}\sim 10^{-33}$ cm, a scale where virtual black holes
continually arise as quantum fluctuations.  It becomes hard to understand why
quantum information can be so readily destroyed at the Planck scale, yet is so
well preserved at the much longer distance scales that we have been able to
explore experimentally --- violations of quantum mechanics, after all, have
never been observed.

Our newly acquired understanding of fault--tolerant quantum computation
provides us with a fresh and potentially fruitful way to think about this
problem.  In Kitaev's spin models, we might imagine that localized processes
that destroy quantum information are quite common.  Yet were we to follow the
evolution of the system with coarser resolution, tracking only the information
encoded in the charges of distantly separated quasiparticles, we would observe
unitary evolution to remarkable accuracy; we would detect no glimmer of the
turmoil beneath the surface.\footnote{Similar language could be used to
characterize the performance of a concatenated code---errors are rare when we
inspect the encoded information with poor resolution, but are seen to be much
more common if we probe the code block at lower levels of concatenation.}

Likewise, it is tempting to speculate that Nature has woven fault tolerance
into her design, shielding the quantum noise at the Planck scale from our view.
The discovery that quantum systems can be stabilized through suitable coding
methods prompts us to ask the question:  Is Nature fault tolerant?  If so, then
quantum mechanics may reign (to excellent accuracy) at intermediate length
scales, but falter both at the Planck scale (where ``errors'' are common) and
at macroscopic scales (where decoherence is rapid).

\section*{Acknowledgments}

This work has been supported in part by DARPA under Grant No. DAAH04-96-1-0386
administered by the Army Research Office, and by the Department of Energy under
Grant
No. DE-FG03-92-ER40701. I am grateful for helpful conversations and
correspondence with Dorit Aharonov, David Beckman, John Cortese, Eric Dennis,
David DiVincenzo, Jarah Evslin, Chris Fuchs, Sham Kakade, Alesha Kitaev, Manny
Knill, Raymond Laflamme, Andrew Landahl, Seth Lloyd, Michael Nielsen, Walt
Ogburn, Peter Shor, Andrew Steane, and Christof Zalka. I especially thank
Daniel Gottesman for many fruitful discussions about fault-tolerant quantum
computation.

\end{document}